\documentclass[useAMS,usenatbib,onecolumn]{mn2e}
\usepackage{graphicx,epsfig,amssymb,amsmath,layout,float,verbatim,rotating,flafter,enumerate,calc}

\def\mnras{MNRAS}
\def\aa{A\&A}
\def\aj{AJ}
\def\apj{ApJ}

\def\cmda{Cel. Mech. Dyn. Astron.}

\title[Appropriate SCF basis sets for orbital studies of galaxies]
{Appropriate SCF basis sets for orbital studies of galaxies and a
`quantum-mechanical' method to compute them}
\author[C. Kalapotharakos, C. Efthymiopoulos \& N. Voglis]{Constantinos Kalapotharakos\thanks{E-mail:ckalapot@phys.uoa.gr (CK);
cefthim@academyofathens.gr (CE)}, Christos
Efthymiopoulos\footnotemark[1] and Nikos Voglis$^\dag$\\
Academy of Athens, Research Center for Astronomy, 4 Soranou
Efesiou Str., GR-11527 Athens, Greece}

\date{Accepted ..........Received .............;in original form ..........}

\pagerange{\pageref{firstpage}--\pageref{lastpage}}

\pubyear{2007}

\begin{document}

\maketitle

\label{firstpage}

\begin{abstract}
We address the question of an appropriate choice of basis
functions for the self-consistent field (SCF) method of simulation
of the $N$-body problem. Our criterion is based on a comparison of
the orbits found in $N$-body realizations of analytical
potential-density models of triaxial galaxies, in which the
potential is fitted by the SCF method using a variety of basis
sets, with those of the original models. Our tests refer to
maximally triaxial Dehnen $\gamma-$models for values of $\gamma$
in the range $0\leq\gamma\leq 1$, i.e. from the harmonic core up
to the weak cusp limit. When an $N$-body realization of a model is
fitted by the SCF method, the choice of radial basis functions
affects significantly the way the potential, forces, or
derivatives of the forces are reproduced, especially in the
central regions of the system. We find that this results in
serious discrepancies in the relative amounts of chaotic versus
regular orbits, or in the distributions of the Lyapunov
characteristic exponents, as found by different basis sets.
Numerical tests include the Clutton-Brock (1973) and the
Hernquist-Ostriker (1992) basis sets, as well as a family of
numerical basis sets which are `close' to the Henquist-Ostriker
basis set (according to a given definition of distance in the
space of basis functions). The family of numerical basis sets is
parametrized in terms of a quantity $\varepsilon$ which appears in
the kernel functions of the Sturm-Liouville equation defining each
basis set. The Hernquist-Ostriker basis set is the $\varepsilon=0$
member of the family. We demonstrate that grid solutions of the
Sturm-Liouville equation yielding numerical basis sets (Weinberg
1999) introduce large errors in the variational equations of
motion. We propose a quantum-mechanical method of solution of the
Sturm-Liouville equation which overcomes these errors. We finally
give criteria for a choice of optimal value of $\varepsilon$ and
calculate the latter as a function of the value of $\gamma$, i.e.,
of the power-law exponent of the radial density profile at the
central regions of the galaxy.
\end{abstract}


\section{Introduction}
The `self-consistent field' (SCF) method of integrating the
gravitational $N$-body problem (Clutton-Brock 1972, 1973; Aoki \&
Iye 1978; Allen, Palmer \& Papaloizou 1990; Hernquist \& Ostriker
1992; Earn and Sellwood 1995; Weinberg 1999) has so far proved
quite useful in the study of particular classes of stellar systems
such as isolated galaxies. A powerful characteristic of this
method is that the gravitational potential, or the force field, at
any point of the particles' configuration space are rendered by
the code in a closed mathematical form, expressed as a series in a
suitable set of basis functions.  The main task of the code is to
calculate the coefficients of a series, the resulting potential of
which (say $V(r,\theta,\phi)$ in spherical coordinates)
corresponds, via Poisson equation $\nabla^2\Phi=4\pi G\rho$, to a
smooth density field $\rho(r,\theta,\phi)$ that is the continuous
limit of the mass distribution of the $N$ particles. For a known
such distribution $\rho(r,\theta,\phi)$, the series coefficients
are given by definite volume integrals of quantities depending on
both $\rho(r,\theta,\phi)$ and on the basis functions. In a
$N$-body simulation, however, we do not have a priori knowledge of
the smooth limit $\rho(r,\theta,\phi)$. Thus, we can only obtain
Monte Carlo estimates of the values of the coefficients based on
the particles' positions, i.e., mass elements
$dm=\rho(r,\theta,\phi)r^2 \sin\theta drd\theta d\phi$ are
replaced in all the volume integrals by point masses (= the
particles) and the integrals are approximated by sums over the
particles' positions. It follows that the overall complexity of
the SCF method is $O(N)$, i.e. linear on the number of particles.
Parallelization is straightforward (Hernquist, Sigurdsson \& Bryan
1995) and applications involving $10^6$ - $10^9$ particles are
tractable on present-day computers.

Owing to its nature, the SCF method is particularly suitable to the
purpose of studying the orbital structure or the global dynamics of
{\it self-consistent} models of galaxies. In general, there are three
ways of producing smooth estimates of the potential of a self-consistent
stellar dynamical system:

a) One chooses an `ad hoc' analytical model for the potential and
then produces a Schwartzchild-type (1979) self-consistent model based
on a large number of orbits. In this case we actually make no estimate
but simply specify the potential in advance. Thus the orbital analysis is
relatively easy (see Efthymiopoulos, Voglis \& Kalapotharakos 2007,
section 4 for an indicative list of references) since it only requires
to integrate orbits in a fixed potential. However, it is well known that
the self-consistent models constructed by Schwartzchild's method are
non-unique, as many different models can be constructed for the same
system which vary in the velocity distribution. Furthermore, the stability
of the Schwartzchild-type models cannot be a priori guaranteed (Merritt
1999; Efthymiopoulos et al. 2007).

b) One makes an $N$-body simulation using one numerical method
(e.g. direct summation, TREE, or mesh code) up to a final time,
e.g., when the system reaches equilibrium. Then one obtains a
smooth estimate of the potential at the final snapshot using a
different method, for example, spline interpolation, fitting with
polynomial or rational functions, grid or the SCF method. When
possible, this method is to be preferred over Schwartzchild's
method since $N$-body equilibria are by definition self-consistent
and also stable. An important early example was given by Sparke
and Sellwood (1987) in the case of a fast rotating barred galaxy,
while more recent examples, in the case of elliptical galaxies,
were given by Contopoulos, Efthymiopoulos \& Voglis 2000; Jesseit,
Naab \& Burkert 2005; Muzzio, Carpintero \& Wachlin 2005 and
Muzzio 2006. Attention should however be paid on that the use of
two different methods to fit the potential, during and after the
$N$-body evolution, may introduce discrepancies in the resulting
analysis of the orbits. The sensitivity of the orbital analysis on
the potential approximation was studied by Carpintero and Wachlin
(2006). These authors found that the relative amount of chaotic
versus regular orbits in a stationary triaxial model depends
significantly on the way chosen to approximate the potential. As
shown in the sequel (sections 3 and 4), this occurs even if two
approximations are similar, e.g., in the case of the SCF method,
two not very different basis sets. This can be due, for example,
to the form of the basis functions, when this does not fit the
morphology of the galaxy, or to errors introduced in the equations
of motion, when a basis set is numerically calculated on a grid
(as suggested by Weinberg 1999). In any case, such numerical
effects affect the orbital analysis in such a way that it becomes
unclear up to what extent the orbital structure found in a
particular system should be attributed to the real dynamics of the
system or to numerical features of the code that fits the
potential.

c) One integrates the $N$-body system as in (b) and then uses {\it
the same} scheme (provided by the $N$-body code) to fit the
potential at the final snapshot after the $N$-body run (Pfenniger
and Frendli 1991, Udry and Martinet 1994, Holley-Bockelmann et al.
2001, 2002; Contopoulos, Voglis \& Kalapotharakos 2002; Voglis,
Kalapotharakos \& Stavropoulos 2002; Athanassoula 2002,
Kalapotharakos, Voglis \& Contopoulos 2004; Shen \& Sellwood 2004;
Kalapotharakos \& Voglis, 2005; Voglis, Stavropoulos \&
Kalapotharakos 2006). This method has internal consistency to a
larger degree than (a) or (b). The SCF method is an obvious
candidate for such simulations. Nevertheless, as already pointed
out, the true `smooth' limit of the distribution of the $N$
particles in an actual $N$-body simulation is not known a priori
(the precise meaning of such a limit has even been questioned in
the literature, Gurzandyan and Savvidy 1986, Kandrup \& Sideris
2003). Thus, the question still remains on how credible can the
reproduction of the orbital structure of an ideally smooth
gravitational system be, when using an SCF method to fit the
potential that is based solely on data of a discretized version of
the system, i.e., of the $N$-body system. A second relevant
question is whether there are means that can be devised so that
this credibility be maximized. In the present paper we deal,
precisely, with the above two questions, focusing our
investigation, as regards in particular the second question, on
the appropriate choice of basis functions for the SCF method and
on an appropriate method to calculate them (when they are not
given analytically).

We propose the following as a numerical test checking the
credibility of a particular SCF method (or of any other potential
fitting method): 1) Choose a pair of smooth potential-density
functions $\rho(\mathbf{r}), \Phi(\mathbf{r})$. 2) Create an
$N$-body realization of the distribution $\rho(\mathbf{r})$ (this
can be done by the Monte-Carlo method). 3) Use the SCF code (or
any other code) in order to calculate a numerical estimate of the
smooth potential $\Phi_{NB}(\mathbf{r})$ from the $N$-body system.
This we call the {\it response} potential. 4) Finally, compare the
orbital structures of the two systems defined by the potentials
$\Phi(\mathbf{r})$, $\Phi_{NB}(\mathbf{r})$. The code can be
characterized as reliable if the two potentials yield a similar
orbital structure. Besides the orbits themselves, a comparison may
involve various quantities characterizing the orbital dynamics
such as, for example, the calculation of the profile of the
forces, frequency analysis of the orbits, Poincar\'{e} phase
plots, Lyapunov characteristic exponents or other indicators of
chaos, etc. Our comparisons in the present refer to the central
profiles of the forces and to the distributions of the Lyapunov
characteristic exponents of the orbits. We also give estimates of
the impact of Poisson noise, introduced at step (2), on the
results. The impact of discreetness was examined in a somewhat
different context by Holley-Bockelmann, Weinberg \& Katz (2005).

We focus on models of elliptical galaxies, in which an important
factor affecting the accuracy of the potential fitting is the
choice of basis functions for the radial part of the potential
expansion (Hernquist \& Ostriker 1992; Hozumi \& Hernquist 1995).
This is related to the fact that the orbits in such systems are
sensitive especially on the radial profile of the forces in the
central parts of the galaxy. On the basis of their central
luminosity profiles, the elliptical galaxies are distinguished in
two groups (Ferarrese et al. 1994; Lauer et al. 1995): a) the
`core' galaxies, with nearly flat or `shallow' surface brightness
profiles, corresponding to power-law density profiles
$\rho(r)\propto r^{-\gamma}$ with $0\leq \gamma\lesssim 1$
(Fridman \& Merritt 1997), and b) the `power-law' galaxies in
which $\gamma\geq 1$, which are further categorized into those
with a `weak cusp' ($\gamma\simeq 1$) or `strong cusp'
(($\gamma\simeq 2$) (Merritt \& Fridman 1996). In case (a) the
force at the centre is equal to zero, while in case (b) the force
is finite, for $\gamma=1$, or infinite, for $\gamma >1$. These
differences in the force field at the centre affect directly the
regular or chaotic character of the orbits (see Efthymiopoulos et
al. 2007 for a review), and imply that a SCF method can only be
successful if it reproduces correctly the central behavior of the
forces. In addition, examples are given in which small numerical
errors in the potential are amplified in the forces, i.e.,
derivatives of the potential, and even more in the {\it
variational equations}, i.e., second derivatives of the potential,
through which the Lyapunov characteristic exponents of the orbits
are evaluated. We find that this may lead to an erroneous
characterization especially of the regular or chaotic character of
the orbits. Furthermore, even if an error appears in only a small
central region, its presence can affect a large number of orbits,
in particular box orbits which pass arbitrarily close to the
centre. Since the box orbits constitute the backbone of many
elliptical galaxies, the dynamical implications of differences in
the central force field are important at least in these galaxies.
This we check by taking as our basic model a maximally triaxial
Dehnen (1993) $\gamma-$model (Merritt \& Fridman 1996) for values
of $\gamma$ in the range $0\leq\gamma\leq 1$, i.e., from the limit
of a harmonic core up to the `weak cusp' limit. Such cases are
characterized by the presence of many box orbits, and indeed, we
find that these orbits are strongly affected by the SCF basis set
used to obtain the response potential, mainly because of
differences in the central force field. This contradicts a claim
by Hernquist \& Ostriker (1992, section 5.2.1) that such
differences are ``probably not significant from a dynamical point
of view''.

In our tests we consider radial basis sets obtained from the
literature, i.e., the Hernquist-Ostriker (1992) and the
Clutton-Brock (1973) basis sets, but also a family of basis sets
computed numerically, as proposed by Weinberg (1999). In computing
the latter, however, we did not use a grid method to solve the
associated Sturm-Liouville boundary value problem (Pruess and
Fulton 1993) because we demonstrate that such methods often result
in large errors appearing in the variational equations of motion
which make use of the second derivatives of a specified basis set
(section 3). Instead, we propose a `quantum-mechanical' method of
solution of the Sturm-Liouville problem which overcomes these
errors. This method is applicable when the Sturm-Liouville
differential equation to be solved is `close' to another
differential equation for which the solution is known analytically
(the definition of distance of two differential operators is given
in section 3). In our examples below we use the
`quantum-mechanical' method in order to calculate numerical basis
sets which are `close' to the Hernquist-Ostriker (1992) basis set,
but improve, however, the representation of the forces at the
central parts of the galaxy. The family is parametrized by a
quantity $\varepsilon$ which appears in the Sturm-Liouville
differential equation (the Hernquist-Ostriker basis set is the
$\varepsilon=0$ member of the family). We then explore which
member of the family better fits, in the $N$-body realization, the
true dynamics of the Dehnen model for a particular value of
$\gamma$. The latter information is given in terms of a function
$\varepsilon(\gamma)$ specifying, essentially, the choice of
optimal basis set as a function of the power-law exponent $\gamma$
of the central density profile of the galaxy (when $\gamma\leq
1$). This information can be used a priori, i.e., one may choose
the optimal basis set for a given $N$-body simulation by measuring
first the value of $\gamma$ (from the $N$-body data).

The paper is organized as follows: section 2 presents the general
formalism of the SCF method in spherical coordinates, following
the same notation as in Weinberg (1999), and then refers to the
appearance of errors due to several previously mentioned sources.
Section 3 describes our `quantum-mechanical' method of
determination of numerical basis sets. Section 4 contains the main
results regarding the choice of an optimal basis set following a
comparison of the orbits in various Dehnen models and in their
respective $N$-body response models. Section 5 summarizes the main
conclusions of the present study.

\section{The Method}

\subsection{Basic Formalism of the SCF method}

We start with the basic formalism of the self-consistent field method in
the case of spherical coordinates, following the same notation as in Weinberg
(1999) for cylindrical coordinates. In the SCF approach, a distribution of
particles in ordinary space is viewed as a Monte Carlo realization of a smooth
density field. This field, which is a continuous and differentiable function
in space and time is given by the integration of the distribution function
$f$ with respect to the velocities:
\begin{equation}\label{rho}
\rho(\mathbf{r},t)= \int_{\mbox{velocities}}
f(\mathbf{r},\mathbf{v},t)d^3\mathbf{v}.
\end{equation}
In the sequel we fix the value of the time $t$ and drop this from the arguments
of $\rho$. In the SCF method the function $\rho(\mathbf{r})$ is expanded in a
truncated series of basis functions in coordinates relevant to the shape of the
system under study. In the case of elliptical galaxies the usual choice are
multipole expansions in spherical coordinates. Let $\rho_{monopole}(r)$ be
the monopole term of the multipole expansion of the density and $\rho_{00}(r)$
a rough estimate that we make of it. Then, we express the monopole term as a
truncated series in terms of radial basis functions $u_{n00}(r)$, i.e.:
\begin{equation}\label{rhomon}
\rho_{monopole}(r) =
\rho_{00}(r)\sum_{n=0}^{n_{max}}b_{n00}u_{n00}(r).
\end{equation}
The use of a discrete spectrum of basis functions $u_{n00}(r)$
(labelled by the `radial quantum number' $n$) follows from
boundary conditions imposed to the system (e.g. finite total mass
and/or finite size). The coefficients $b_{n00}$ are unknown and
the main task of the $N$-body code is to specify their values.
Eq.(\ref{rhomon}) reflects our expectation that a linear
combination of the functions $u_{n00}(r)$ can fit the residuals of
the true monopole term of the real density with respect to our
initial estimate $\rho_{00}(r)$ which acts as an envelope in front
of the sum in the r.h.s.. In reality, the fitting is efficient if
only a small number of terms are needed in (\ref{rhomon}) (the
uppermost limit of $n$, imposed by the $N$-body resolution is
$n_{max}=O(N^{1/3})$ but in practice we use a number of terms
which is one order of magnitude smaller than this limit (Palmer
1994)). This, on its turn, depends crucially on the initial
estimate $\rho_{00}(r)$, which appears not only directly in
Eq.(\ref{rhomon}) but also, as shown below, indirectly, through
the Sturm-Liouville differential equation which specifies the
basis functions $u_{n00}(r)$. At any rate, in the same way as for
the monopole term, we can make in advance some estimate of the
profile of the multipole terms of the density by choosing estimate
functions $\rho_{lm}(r)$ and then fit the residuals of the true
multipole terms with respect to the functions $\rho_{lm}(r)$ via
series in respective basis functions $u_{nlm}(r)$. That is, the
density is finally written as:
\begin{equation}\label{rhoscf}
\rho(r,\theta,\phi) = \sum_{l=0}^{l_{max}}
\sum_{m=-l}^l\sum_{n=0}^{n_{max}}b_{nlm}\rho_{lm}(r)u_{nlm}(r)
Y_l^m(\theta,\phi)
\end{equation}
where $Y_l^m(\theta,\phi)$ are spherical harmonics (e.g. Binney
and Tremaine 1987, pp.655-656). Similarly as for the monopole
term, the coefficients $b_{nlm}$ are unknown while the functions
$u_{nlm}(r)$ are specified in advance via solutions of a
Sturm-Liouville differential equation in which the functions
$\rho_{lm}(r)$ also appear.

The determination of the basis set of functions $u_{nlm}(r)$ is
done as follows: Repeating for the gravitational potential the
same procedure as for the density, i.e., selecting in advance some
estimate functions $\Phi_{lm}(r)$ for the profiles of the various
multipole potential terms we also write the potential as
\begin{equation}\label{phiscf}
\Phi(r,\theta,\phi) = \sum_{l=0}^{l_{max}}
\sum_{m=-l}^l\sum_{n=0}^{n_{max}}c_{nlm}\Phi_{lm}(r)u_{nlm}(r)
Y_l^m(\theta,\phi)
\end{equation}
with unknown coefficients $c_{nlm}$, and then match
Eqs.(\ref{rhoscf}) and (\ref{phiscf}) via Poisson equation. This
yields finally (in units in which $G=1$):
\begin{equation}\label{stlv1}
-\dfrac{d}{dr}\left(r^2 \Phi_{lm}^2
\dfrac{du_{nlm}}{dr}\right)+\left[l(l+1)\Phi_{lm}^2-
\Phi_{lm}\dfrac{d}{dr}\left(r^2\dfrac{d\Phi_{lm}}{dr}\right)\right]u_{nlm}=
-(4\pi \lambda_{nlm} r^2 \Phi_{lm}\rho_{lm})u_{nlm}
\end{equation}
with $\lambda_{nlm} = b_{nlm}/c_{nlm}$. Equation (\ref{stlv1}),
supplemented with appropriate boundary conditions, is a case of
the Sturm-Liouville eigenvalue problem
\begin{equation}\label{stlv2}
-\dfrac{d}{dr}\left[p_{lm}(r)\dfrac{du}{dr}\right]+q_{lm}(r)u=\lambda
w_{lm}(r)u
\end{equation}
with
\begin{subequations}
\begin{align}\label{pqwsl}
p_{lm}(r)&=r^2 \Phi_{lm}^2(r)\\
q_{lm}(r)&=l(l+1)\Phi_{lm}^2(r)-\Phi_{lm}(r)\frac{d}{dr}\left(r^2
\frac{d\Phi_{lm}(r)}{dr}\right)\\
w_{lm}(r)&=-4\pi r^2 \Phi_{lm}(r)\rho_{lm}(r).
\end{align}
\end{subequations}
From the form  of (\ref{stlv1}) it follows that the functions
$u_{nlm}(r)$ are given by the eigenfunctions of a Sturm-Liouville
differential operator
\begin{equation}\label{stlvop}
{\cal
L}_{lm}=-\frac{d}{dr}\left[p_{lm}(r)\frac{d}{dr}\right]+q_{lm}(r)
\end{equation}
acting on functions belonging to a Hilbert space with the inner product
definition
\begin{equation}\label{inpro}
<f|g> = \int_{r_a}^{r_b} f(r)g(r)w_{lm}(r)dr~~
\end{equation}
where $r_a,r_b$ are two radii at which boundary conditions must be given in
the form (Pruess \& Fulton 1993)
\begin{subequations}
\label{bcstlv}
\begin{align}
a_1 u-a_2\left(p_{lm}(r) \frac{du}{dr}\right)&=\lambda_{nlm}
\left(a_1' u-a_2'
\frac{du}{dr}\right)~~~~\mbox{at}~r=r_a\\
b_1 u+b_2\left(p_{lm}(r)
\frac{du}{dr}\right)&=0~~~~~~~~~~~~~~~~~~~~~~~~~~~~~~\mbox{at}~r=r_b
\end{align}
\end{subequations}
with constants $a_1,a_2,a_1',a_2',b_1,b_2$. The problem admits a
discrete set of solutions, i.e., a discrete set of eigenvalues
$\lambda_{nlm}$ and eigenvectors $u_{nlm}$, $n=0,1,2,...$ of
${\cal L}_{lm}$. In this way, the basis functions $u_{nlm}(r)$ are
specified by the initial choice of estimate functions
$\rho_{lm}(r),\Phi_{lm}(r)$, which are hereafter called the kernel
functions of the Sturm-Liouville problem,  and by the boundary
conditions. The index $n$ is called the radial quantum number.

In galactic problems, the radii $r_a, r_b$ are usually set equal
to $r_a=0$ (centre of the system), and $r_b=R_p$ or
$r_b\rightarrow\infty$, depending on whether we consider a system
ending at a finite radius $R_p$ or at infinity. In the former
case, the boundary conditions at $r_b=R_p$ represent the request
of continuity, and of continuous derivative, of the potential
function at the point $R_p$ where we pass from Poisson to Laplace
equation. Such boundary conditions can always be cast in the form
(\ref{bcstlv}).

If the kernel functions $\Phi_{lm}(r)$, $\rho_{lm}(r)$ satisfy
Poisson equation $\nabla^2\Phi_{lm}(r)=4\pi G\rho_{lm}(r)$, then
they are called a {\it potential-density pair} of functions. In
that case we always have $\lambda_{0lm}=1$ and $u_{0lm}=const$.
Independently of whether the kernel functions are
potential-density pairs or not, the eigenfunctions $u_{nlm}(r)$ of
the Sturm-Liouville problem (\ref{stlv1}) are always orthogonal
with respect to the inner product definition (\ref{inpro}).

In an $N$-body simulation we use the above formalism in order to
obtain a smooth potential function as follows:

a) We specify in advance the sets $\rho_{lm}(r)$, $\Phi_{lm}(r)$,
and then $u_{nlm}(r), \lambda_{nlm}$, as explained above.

b) Given the particles' positions, we find estimates of the
coefficients $b_{nlm}$ of an underlying `smooth' density field, by
exploiting the orthogonality relation
$<u_{nlm}|u_{n'lm}>=\delta_{n,n'}$, together with the
orthogonality of the spherical harmonic functions. Namely,
equation (\ref{rhoscf}) is inverted, yielding the value of any
particular coefficient $b_{nlm}$ as
\begin{equation}\label{bnml1}
b_{nlm} = \int_{r_a}^{r_b}\int_0^\pi\int_0^{2\pi} R(r,\theta,\phi)
dr d\theta d\phi
\end{equation}
where
$$
R(r,\theta,\phi)=-4\pi\Phi_{lm}(r)u_{nlm}(r)Y_l^{m*}(\theta,\phi)
\rho(r,\theta,\phi)r^2\sin\theta.
$$
Assuming now that the positions of the $N$ particles provide a
discrete realization of the smooth density field
$\rho(r,\theta,\phi)$, the volume integral (\ref{bnml1}) can be
evaluated by the Monte Carlo method, i.e., as a sum over the
particles' positions:
\begin{equation}\label{bnml2}
b_{nlm} \simeq \sum_{i=1}^N
-4\pi\Phi_{lm}(r_i)u_{nlm}(r_i)Y_l^{m*}(\theta_i,\phi_i).
\end{equation}

c) We finally calculate the coefficients of the potential series
as $c_{nlm}=b_{nlm}/\lambda_{nlm}$.

The use of a discrete sum instead of a continuous integration
implies that, contrary to what the term `smooth field' might
suggest, there is always some numerical noise in the system
introduced by discreteness effects, which produces `relaxation'
effects in the $N$-body simulation (see Palmer 1994, or Weinberg
1996 for a detailed discussion). In our numerical examples in the
sequel we examine in detail the effect of this type of noise on
the orbits. At any rate, as already mentioned the goodness of the
fit of the smooth potential of a system by the SCF series depends
crucially on the choice of basis functions $u_{nlm}$ which, on
their turn, depend on the choice of appropriate kernel functions
$\rho_{lm}(r)$, $\Phi_{lm}(r)$. To this we now turn our attention.

\subsection{The choice of basis functions}

In order to determine a suitable basis set, a common strategy in
the literature is to use kernel functions $\rho_{lm}(r)$,
$\Phi_{lm}(r)$ such that the solutions of (\ref{stlv1}) are
reduced to some simple form given, e.g., in terms of special
functions or in a closed polynomial form. We mention the following
examples (see Hernquist \& Ostriker 1992 for a detailed review):

\textbf{(i)} the Clutton-Brock (1973, hereafter CB) set. The
kernel functions are:
\begin{subequations}
\begin{align}
\label{CBa}
\rho_{lm}(r)&=\frac{\sqrt{4\pi}}{4\pi}(2\,l+1)(2\,l+3)\dfrac{a^2\,r^l}{(a^2+r^2)^{l+5/2}}\\
\label{CBb}
\Phi_{lm}(r)&=-\sqrt{4\pi}\dfrac{r^l}{(a^2+r^2)^{l+1/2}}
\end{align}
\end{subequations}
with boundary conditions
$$
\left. \dfrac{du_{nlm}(r)}{dr}\right|_{r=0}=0 \text{~~and~~}
\lim_{r\rightarrow\infty} \dfrac{du_{nlm}(r)}{dr}=0.
$$
In the form (\ref{bcstlv}) these conditions are given e.g. as
$a_1=a_1'=a_2'=0$, $a_2=1$, and $b_1=0$, $b_2=1$. The zeroth order
term of the density expansion
$$
\rho_{000}=\frac{3}{\sqrt{4\pi}}\frac{a^2}{(a^2+r^2)^{5/2}}
$$
yields a half-mass radius equal to unity, $r_{1/2}=1$, if $a\simeq 0.77\simeq
10/13$. The corresponding potential is
$$
\Phi_{000}=-\sqrt{4\pi}\frac{1}{(a^2+r^2)^{1/2}}.
$$
The above set $\rho_{000},\Phi_{000}$ is the well known Plummer
model. The eigenfunctions $u_{nlm}(r)$ are Gegenbauer polynomials
of the form $C_n^{l+1}(\xi)$ (independent of $m$) where
$\xi=\dfrac{r^2-a^2}{r^2+a^2}$.

\textbf{(ii)} the set of Allen et al. (1990). In this case there
are no kernel functions of the potential or density, i.e.
\begin{equation}\label{pal}
\rho_{lm}(r)=1,~~~\Phi_{lm}(r)=-1.
\end{equation}
The inner boundary condition is $du_{nlm}(0)/dr=0$ while the outer
boundary condition corresponds to the matching of the solutions of
the Poisson and Laplace equations at some finite radius $r_b=R_p$
$$
\left.\dfrac{du_{nlm}(r)}{dr}\right|_{r=0} =
0,~~~\left.r\dfrac{du_{nlm}(r)}{dr}+(l+1)u_{nlm}(r)\right|_{r=R_p}=0.
$$
In the form (\ref{bcstlv}) we set $a_1=a_1'=a_2'=0$, $a_2=1$, and
$b_1=l+1$, $b_2=1/R_p$. The eigenfunctions $u_{nlm}(r)$ are
spherical Bessel functions. This set better fits systems with a
harmonic core.

\textbf{(iii)} the Hernquist-Ostriker (1992, hereafter HO) set.
The kernel functions are
\begin{subequations}
\begin{align}
\label{hoa}
\rho_{lm}(r)&=\sqrt{4\pi}\dfrac{1}{2\pi}\dfrac{(2l+1)(l+1)}{r}
\dfrac{a\,r^l}{(a+r)^{2l+3}}\\
\label{hob} \Phi_{lm}(r)&=-\sqrt{4\pi}\dfrac{r^l}{(a+r)^{2l+1}}.
\end{align}
\end{subequations}
The zeroth order term of the density
$$
\rho_{000}(r)=\sqrt{4\pi}\frac{1}{2\pi}\frac{1}{r}
\frac{a}{(a+r)^{3}}
$$
yields a half-mass radius equal to unity, $r_{1/2}=1$, if
$a=(1+\sqrt{2})^{-1/2}$ which is our choice of value of $a$ in the
sequel. The boundary conditions are
$\dfrac{du_{nlm}}{dr}=\mbox{finite}$ at $r_a=0$ and the same as in
the CB set at infinity. In the form (\ref{bcstlv}) we give
precisely the same constants as in the CB set. The eigenfunctions
$u_{nlm}$ are Gegenbauer polynomials of the form
$C_n^{2l+\frac{3}{2}}(\xi)$ where $\xi=\dfrac{r-a}{r+a}$.

The choice of an appropriate set of basis functions is related to
the morphological characteristics of the system under study. For
example, if the kernel function $\rho_{00}(r)$ is selected as in
the HO basis set (Eq.(\ref{hoa})), this function yields a
power-law cusp $\rho_{00}(r)\sim r^{-1}$at the centre, so the HO
set cannot easily fit systems in which the density cusp at the
centre is shallower than this law. In fact, Hernquist \& Ostriker
(1992, subsection 2.3) give an idealized example in which a flat
profile at the centre can be represented by a linear combination
of the $n=0$ and $n=1$ monopole terms of the HO basis set, because
the coefficients of these terms are balanced in a way so as to
eliminate the power-law dependence $r^{-1}$. However, we can show
that if, due, for example, to the Poisson noise, the balance is
slightly distorted in an actual $N$-body calculation, the $r^{-1}$
singular behavior of the density at the centre reappears as a
result of the numerical error. In our normalization units and
constants ($a=(1+\sqrt{2})^{-1/2}$, which corresponds to $a=1$ in
Hernquist \& Ostriker 1992), the example refers to the spherical
density profile:
\begin{equation}\label{rhofnt}
\rho_0(r)={3\over 4\pi}{a\over (a+r)^4}
\end{equation}
which is clearly flat at the centre. While the kernel function
$\rho_{00}(r)$ given by Eq.(\ref{hoa}) goes as $\rho_{00}(r)\sim
r^{-1}$ at the centre, a combination of the two first monopole
terms of the radial basis set allows one to remove the singularity
and obtain the profile (\ref{rhofnt}). Precisely, we have:
\begin{equation}\label{hobasis}
u_{nlm}={a^{l+{1\over 2}}2^{4l+3}\Gamma(2l+{3\over 2})\over 4\pi}
\sqrt{n!(n+2l+{3\over 2})\over (1+2l)(l+1)\Gamma(n+4l+3)}
C_n^{2l+\frac{3}{2}}\left(\frac{r-a}{r+a}\right)
\end{equation}
so that
$$
\rho_{000}={\sqrt{3}a^{3/2}\over 2\pi r(a+r)^3}
$$
and
$$
\rho_{100}={\sqrt{15}a^{3/2}(r-a)\over 2\pi r(a+r)^4}.
$$
Thus
$$
\rho_{000}(r)+c \rho_{100}(r)={\sqrt{3}a^{3/2}\over 2\pi r(a+r)^3}
\Bigg[1+{c\sqrt{5}(r-a)\over r+a}\Bigg]
$$
and, if $c=1/\sqrt{5}$ (which corresponds to $c=1/12$ in the units
of Hernquist \& Ostriker (1992)), we find
\begin{equation}\label{rhofntho}
\rho_{000}+{1\over \sqrt{5}}\rho_{100}= {\sqrt{3}a^{3/2}\over \pi
(a+r)^4}\propto \rho_0(r).
\end{equation}
However, the balance in Eq.(\ref{rhofntho}) is quite sensitive to
numerical errors. For example, if, due to discreteness effects,
there is a small error $\epsilon_0$ in the ratio $c$ of the
numerical coefficients of $\rho_{000}$ and $\rho_{100}$, i.e.,
$c=1/\sqrt{5}+\epsilon_0$, the error introduced in the evaluation
of the density is
\begin{equation}\label{rhofntho2}
\rho_{000}+({1\over
\sqrt{5}}+\epsilon_0)\rho_{100}= \rho'_0(r)={\sqrt{3}a^{3/2}\over \pi (a+r)^4}
+{1\over r}\Bigg({3\epsilon_0(r-a)\over 8\pi (a+r)^4}\Bigg)
\end{equation}
and we have $\lim_{r\rightarrow 0}\rho'_0(r)=\infty$, i.e. the
density as calculated by the SCF method with the HO basis set
becomes singular at the centre.

Fig.~1 shows a numerical calculation of the above effect. Using a
$(50\times 100\times 200)$ spherical polar grid in
$(\theta,\phi,r)$ respectively, logarithmic in the radii (from
$r=0$ to $r=15$), and linear in the angles, we produced an
$N$-body realization of the spherical system (\ref{rhofnt}) with
$N=1.25 \times 10^7$ particles as follows: a) a theoretical value
of the number of particles in each grid cell was calculated, that
is given by $\Delta N_{th} = \rho_0(r_c)r_c^2\sin\theta_c\Delta r
\Delta\theta \Delta\phi$, where $r_c,\theta_c$ are the values of
$r$ and $\theta$ at the centre of the cell. b) A random integer
number $\Delta N$ is then determined which has a Poisson
distribution corresponding to a mean value $<\Delta N>=\Delta
N_{th}$ and dispersion $\sigma=\Delta N_{th}^{1/2}$. c) The values
$\Delta N$ for all the cells were normalized so that the total
number of particles is kept equal to $N$. d) Each cell was filled
with $\Delta N$ particles located randomly within the cell with a
uniform distribution. Finally, we used this distribution of
particles to calculate the coefficients of the ten first monopole
terms $b_{n00}$, $n=0,\ldots,9$, of the sum (\ref{rhomon}) with
the HO basis set.

\begin{figure}
\centerline{\includegraphics[width=15cm]{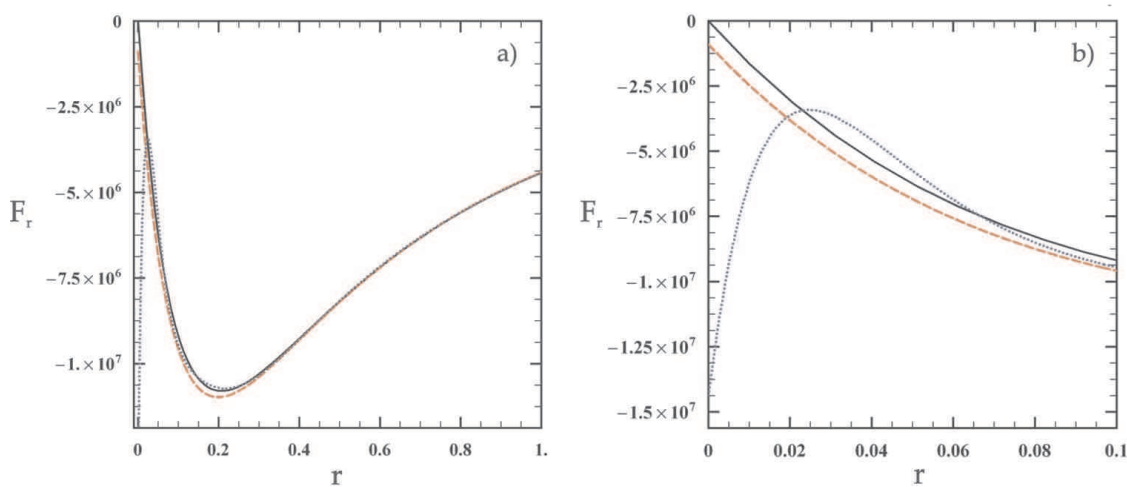}}
\caption{\textbf{(a)} The profile of the radial force $F_r(r)$
corresponding to a mass distribution with the density profile
(\ref{rhofnt}) (with $1.25\times 10^7$ particles) as derived
analytically (solid curve) and via the SCF calculation by the
first two monopole functions ($n=0,1$, $l=m=0$) (dashed red curve)
and by the first ten monopole functions ($n=0-9$, $l=m=0$) (dotted
blue curve) of the HO radial basis set implemented on an $N$-body
realization of this mass distribution (see text for details).
\textbf{(b)} A zoom of (a) in the central region of the model. For
large radii the force is represented very good in both cases (two
and ten monopole terms expansion). For small radii the first two
monopole terms yield a small but definitely finite value of the
force at $r=0$ while the first ten monopole terms yield a
conspicuous error in the force near the centre $r=0$.}
\end{figure}
The first two coefficients in the above simulation obtain values
yielding a ratio $b_{000}/b_{100} =2.27381$, which is close to,
but not exactly equal to the theoretical ratio $b_{000}/b_{100}
=\sqrt{5}=2.236067$. In this case the relative error of the two
coefficients is due to the fact that the $N$-body system is
truncated at a radius $r=15$ after which the density field
(\ref{rhofnt}), for the used number of particles and grid, is
seriously undersampled (e.g. we have less than 1 particles per
cell). This error in the sampling can be partly reduced by
introducing variable masses of the particles (Sigurdsson,
Hernquist \& Quinlan 1995) so that there are more particles
tracing the smooth mass distribution at the outer radii. Even so,
however, the error cannot be reduced further from the Poisson
noise limit caused by the discrete number of particles. This is
estimated as follows: We evaluate the integral (Eq.(\ref{bnml1})
by truncating the radius at $r_b=15$ (instead of infinity) and
find theoretical coefficients $b_{000,r_b=15}^{theoretical}$,
$b_{100,r_b=15}^{theoretical}$ from the theoretical density
$\rho(r,\theta,\phi)\equiv\rho_0(r)$, and also numerical
coefficients by the sums (\ref{bnml2}) yielding a relative error
$$
\left|\frac{\Delta
b_{000,r_b=15}}{b_{000,r_b=15}^{theoretical}}\right|\simeq
2.7\times 10^{-4},~~ \left|\frac{\Delta
b_{100,r_b=15}}{b_{100,r_b=15}^{theoretical}}\right|\simeq
8.4\times 10^{-4}.
$$
Both these relative errors are comparable to a Poisson noise error
of order $O(1/\sqrt{N})$, with $N\sim 10^7$.

Now, both errors due to the truncation and to the Poisson noise
contribute in that, while the overall fitting of the radial force
profile by the HO basis set with the two first terms ($n=0,1$) is
quite satisfactory (Fig.~1a), the force obtained at the centre by
the same terms is finite (Fig.~1b). Even so, one may argue that
the finite value of the force is rather small. However, this only
happens because we considered the first two terms in the radial
expansion, while, in a typical $N$-body simulation in which we are
not aware of whether the density profile is close to some
idealized example, we typically use many more terms (say up to
$n=9$). Theoretically, the coefficients of these terms in
Eq.(\ref{rhofntho}) are equal to zero. In practice, however, we
find that these terms also have small non-zero values which,
because of the $1/r$ singularity in the HO kernel function
$\rho_{00}(r)$, increase dramatically the error in the central
force (Fig.~1b). Furthermore, while the appearance of the error
can in principle be reduced to very small values of $r$, it always
affects orbits which pass arbitrarily close to the centre, such as
the box orbits. This, and other numerical effects, will be
examined in the subsequent sections.

\section{A `quantum mechanical' method of determination
of numerical basis sets}

\begin{figure}
\centerline{\includegraphics[width=\textwidth]{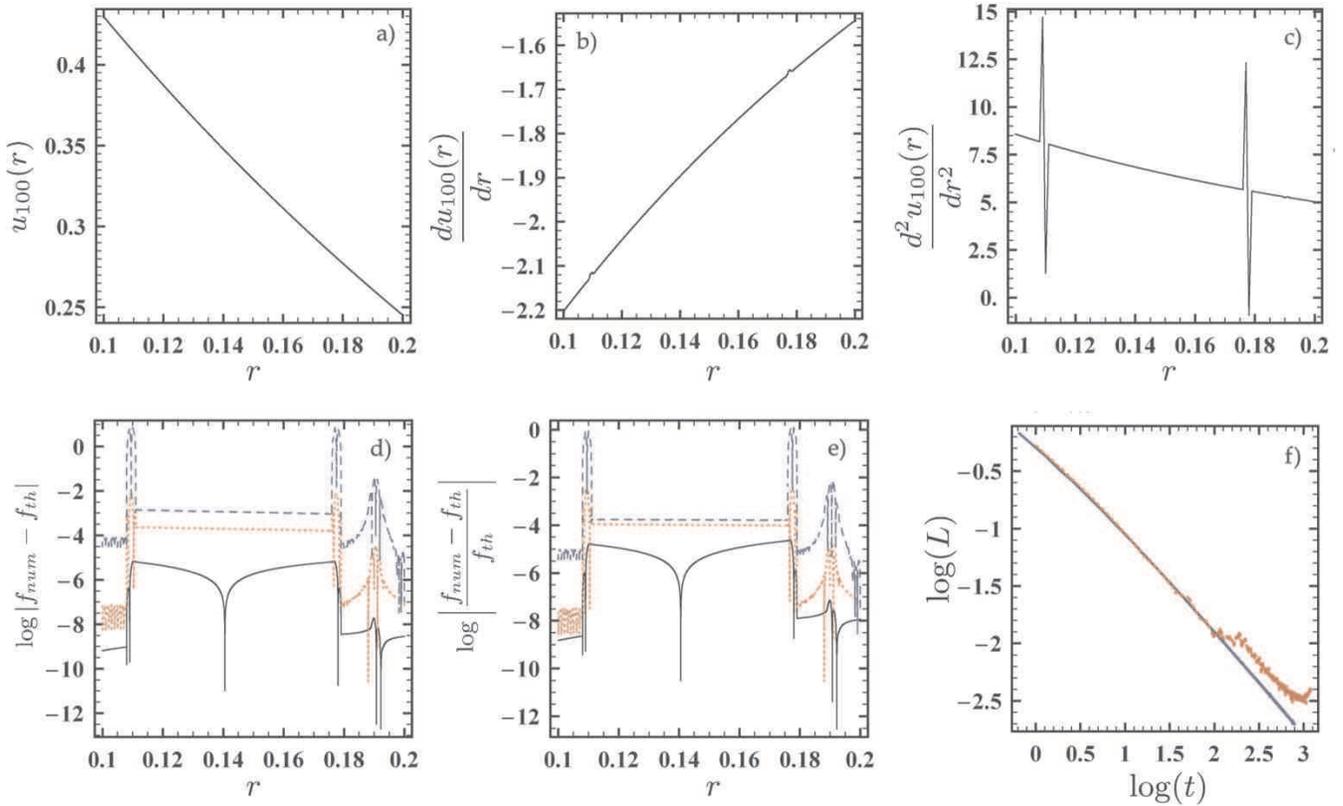}}
\caption{\textbf{(a)} The HO eigenfunction $u_{100}(r)$ $(n=1,
l=m=0)$ versus $r$ as derived by the SLEDGE solver of the
corresponding Sturm-Liouville problem, after a cubic spline
interpolation of the numerical values on a grid of stepsize
$dr=0.001$. \textbf{(b)} The derivative of the curve (a) as
calculated by the interpolating function. Two non smooth spots are
just visible at $r\simeq 0.11$ and $r\simeq 0.175$ respectively.
\textbf{(c)} Second derivative of (a). The error at the two spots
is now clearly amplified, creating large errors in the variational
equations of motion in which the second derivatives of the
eigenfunction appear. \textbf{(d)}, \textbf{(e)} The absolute and
relative errors of the interpolated numerical solution with
respect to the analytical solution in logarithmic scale for the
function $u_{100}$ (black solid line), and for the derivatives
$du_{100}(r)/dr$ (red dotted line) and $d^2u(r)_{100}(r)/dr^2$
(blue dashed line). The error in the first derivative is amplified
with respect to the error in the eigenfunction, while the error in
the second derivative becomes of order unity at the spots.
\textbf{(f)} The evolution of the finite time Lyapunov
characteristic number (LCN) for one orbit in the maximally
triaxial Dehnen model with $\gamma=0.4$ (see section 4), initial
conditions $x_0\simeq 0.045760$, $y_0\simeq 0.004794$, $z_0\simeq
0.001598$, $v_{x0}=0$, $v_{y0}=0$, $v_{z0}=0$, when the potential
is reproduced by the HO basis set calculated analytically (blue
solid curve) or numerically (by the SLEDGE code) on a grid (red
dashed curve). The expansions in both cases reach up to
$n_{max}=9$ and $l_{max}=4$. The LCN curve shows the orbit to be
regular in the first case, and chaotic in the second case with
$LCN\approx 10^{-2.5}$. }
\end{figure}

Weinberg (1999) stressed the need for an adaptive algorithm
producing basis sets tailored to the morphological details of the
system under study, so that the series representation of the
potential has good convergence properties and numerical
instabilities such as that mentioned in the previous section are
avoided. Translated to spherical coordinates, Weinberg's proposal
is essentially the following:

a) Choose kernel functions $\rho_{lm}(r),\Phi_{lm}(r)$ which are
as good estimators as possible of the profiles of the $N$-body
system to be run.

b) Solve numerically the Sturm-Liouville boundary value problem
(\ref{stlv2}) by special solvers such as the SLEDGE code (Pruess
\& Fulton 1993). Such solvers are based on variants of the
shooting method, and they yield the solutions $u_{nlm}$ in
tabulated form, i.e., at the points of a grid.

In principle, Weinberg's proposal extends considerably the
applicability of the SCF method by enlarging the freedom of choice
of kernel functions $\Phi_{lm}(r), \rho_{lm}(r)$, which can even
be adaptive, i.e., change in the course of an $N$-body simulation.
However, we show now that the use of tabulated values on a grid
imposes restrictions to accuracy so that we have to devise an
alternative method for the solution of the Sturm-Liouville
problem.

To this end, we point out first that when the functions $u_{nlm}$
are known only on a grid, the first and second derivatives of the
potential can be calculated directly only by a finite difference
scheme. However, the use of any such scheme whatsoever cancels the
property of smoothness of the potential. In order to retain the
latter we can use an interpolating function for each tabulated
function $u_{nlm}$, with sufficient number of continuous
derivatives, that joins the values of any function $u_{nlm}(r)$ at
successive points of the radial grid. However, if there are small
numerical errors at the grid points, these errors also show up
when one calculates the derivatives either by finite differencing
or by the derivatives of the interpolating function. Fig.~2 shows
how does the error appear when using interpolation. The SLEDGE
solver was used in order to produce numerically the HO basis set
$u_{nlm}$, which is given analytically by Eq.(\ref{hobasis}) with
$l=0,\ldots,4$ and $n=0,\ldots,9$, so that comparisons between the
numerical and analytical solution can be made. The solution was
required on a grid of $2.2\times 10^4$ points in the interval
$0\leq r\leq 22$, with a tolerance $10^{-8}$. After the basis
functions were calculated, each function was interpolated by a
cubic spline, so as to secure the continuity of the first and
second derivative of the interpolating function at the grid
points. Fig.~2a shows the interpolated numerical solution in the
case $n=1$ and in the interval $0.1\leq r\leq 0.2$. Although
SLEDGE could not reach the requested input tolerance $10^{-8}$ in
many parts of the solution, the error actually rendered by SLEDGE
in the values of the function $u_{100}$ was still relatively small
(of order $10^{-6}$ or below, Fig.~2d,e). However, this small
error is amplified when one calculates the derivatives
$\dfrac{du_{100}}{dr}$ (Fig.~2b) via the interpolating function.
Then the error becomes, in general, of order $10^{-4}$, reaching
up to $10^{-2}$ at some particular points of the numerical
solution (Fig.~2d,e). In fact, a careful examination of these
points showed that the grid solution had small jumps there, of
order $10^{-6}$, and these were forcing the interpolating function
to introduce artificial points at which the convexity was
seriously distorted (a function could even turn locally from
convex to concave). Thus, when calculating the second derivative
$\dfrac{d^2u_{100}}{dr^2}$, the error was dramatically amplified
at these particular points (Fig.~2c). We see that the error
locally becomes of order unity, while the overall error in the
second derivatives ranges from $10^{-4}$ to $10^{-2}$, i.e.,
similarly as for the forces. We found similar numerical effects in
practically all the functions $u_{nlm}$ that were determined
numerically by the SLEDGE solver. Furthermore, when orbits were
calculated in an $N$-body realization of a maximally triaxial
Dehnen $\gamma-$model, for $\gamma=0.4$ (see next section), using
the HO basis set calculated numerically up to $n_{max}=9$,
$l_{max}=4$, we found orbits that were affected by these errors to
such an extent that artificial positive Lyapunov characteristic
numbers were introduced in cases of orbits that were in fact
regular (Fig.~2f).

In order to overcome the above numerical difficulties, we suggest
now a `quantum-mechanical' method of solution of the
Sturm-Liouville problem (\ref{stlv2}) which is similar to the
quantum-mechanical perturbation theory of bound eigenstates (e.g.
Merzbacher, 1961, pp.413-437). This method is applicable when a
numerical basis set is required that is not very different from
another basis set known analytically . Let $u^{(0)}_{nlm}(r)$ be
the functions of the known set. These are solutions of the problem
(\ref{stlv2}) for a Sturm-Liouville operator ${\cal
L}^{(0)}_{lm}$, which is determined by a choice of kernel
functions as, e.g., in subsection 2.2, and for specific boundary
conditions of the form (\ref{bcstlv}). Let, now, ${\cal L}_{lm}$
be a different Sturm-Liouville operator corresponding to a
different choice of kernel functions. We seek to solve the
eigenfunction problem
\begin{equation}\label{stlv3}
{\cal L}_{lm} u_{lm}(r) = \lambda w_{lm}(r)u_{lm}(r)
\end{equation}
i.e., find the spectrum of successive eigenvalues $\lambda_{n'}$
and eigenvectors $u_{n'lm}(r)$ of ${\cal L}_{lm}$, $n'=0,1,...$,
for which we request to satisfy the same boundary conditions as
those satisfied by the functions $u^{(0)}_{nlm}(r)$. Since the
latter form a complete basis of the space of functions with the
given boundary conditions, any function can be written as a linear
combination of them. We thus write
\begin{equation}\label{uu0}
u_{n'lm}(r)=\sum_{n=0}^{\infty}d_{n',nlm}u^{(0)}_{nlm}(r).
\end{equation}
In view of (\ref{stlv3}), Eq.(\ref{uu0}) takes the form
\begin{equation}\label{stuu0}
\sum_{n=0}^{\infty}d_{n',nlm}{\cal L}_{lm}u^{(0)}_{nlm}(r)=
\lambda_{n'}\sum_{n=0}^{\infty}d_{n',nlm}w_{lm}(r)u^{(0)}_{nlm}(r).
\end{equation}
Multiplying both sides of (\ref{stuu0}) with
$\dfrac{w^{(0)}_{lm}(r)u^{(0)}_{jlm}(r)}{w_{lm}(r)}$, integrating
over all radii from $r_a$ to $r_b$, and recalling the
orthogonality of the functions $u^{(0)}_{nlm}(r)$ yields the
linear set of equations:
\begin{equation}\label{stuu02}
\sum_{n=0}^{\infty}<u^{(0)}_{jlm}|{\cal
M}_{lm}|u^{(0)}_{nlm}>d_{n',nlm} =\lambda_{n'} d_{n',jlm}
\end{equation}
where the linear operator ${\cal M}_{lm}$ is defined as
\begin{equation}\label{calm}
{\cal M}_{lm} = {w^{(0)}_{lm}(r)\over w_{lm}(r)}{\cal L}_{lm}
\end{equation}
and
\begin{equation}\label{calmin}
<u^{(0)}_{jlm}|{\cal M}_{lm}|u^{(0)}_{nlm}>\equiv
\int_{r_a}^{r_b}u^{(0)}_{jlm}(r){\cal M}_{lm}u^{(0)}_{nlm}(r)dr.
\end{equation}
The problem (\ref{calmin}) is equivalent to the original eigenvalue problem
and it can be written in the matrix form:
\begin{equation}\label{calmin2}
{\cal H}_{lm}\cdot{\mathbf d_{n',lm}}=\lambda{\mathbf d_{n',lm}}
\end{equation}
where ${\mathbf d_{n',lm}}$ is a column vector with entries equal
to the coefficients $d_{n'nlm}$, that is ${\mathbf
d}_{n',lm}^T=(d_{n',0lm},d_{n',1lm} ,d_{n',2lm},\ldots)$, and
${\cal H}_{lm}$ is a matrix with entries ${\cal
H}_{lm,ij}=<u^{(0)}_{ilm}|{\cal M}_{lm}|u^{(0)}_{jlm}>$. This
formally corresponds to what is referred to as the `Hamiltonian
matrix' in quantum mechanics, while Eq.(\ref{calmin2}) is
analogous to the quantum mechanical procedure of diagonalization
of the Hamiltonian matrix.

\begin{figure}
\centerline{\includegraphics[width=14cm]{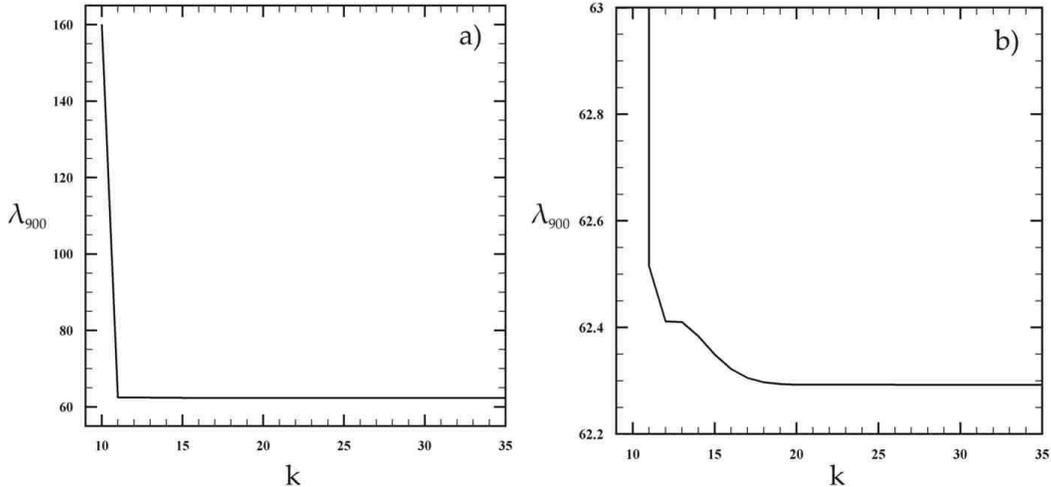}}
\caption{\textbf{(a)} The numerical value of the eigenvalue
$\lambda_{900}$ of the new basis set as a function of the rank $k$
(order of truncation of the matrix $\cal H$) when
$\varepsilon=0.02$. Note that we need at least a $10\times 10$
matrix in order to start calculating this eigenvalue. \textbf{(b)}
A zoom of (a) in which the convergence of the eigenvalue
$\lambda_{900}$ to a final value, as $k$ increases, is shown in
greater detail. The final value is practically reached when
$k\gtrsim 20$.}
\end{figure}

The numerical steps in order to solve the Sturm-Liouville problem (\ref{stlv3})
are then summarized as follows:

a) Starting from a known basis set $u^{(0)}_{nlm}$, calculate the
integrals (\ref{calmin}) and hence the matrix ${\cal H}_{lm}$. In
fact, the matrix ${\cal H}_{lm}$ contains an infinite number of
entries, thus we can only compute $k\times k$ truncations of
${\cal H}_{lm}$. Nevertheless, the determination of the
eigenvalues and eigenvectors of ({\ref{calmin2}) for different
truncation orders $k$ is a convergent procedure (numerical
examples are given below). The convergence is faster when the set
of eigenfunctions $u^{(0)}_{nlm}$ and $u_{nlm}$ correspond to
`nearby' operators ${\cal M}_{lm}^{(0)},{\cal M}_{lm}$. By this we
mean that the kernel functions $\Phi^{(0)}_{lm}(r),\Phi_{lm}(r)$,
and $w^{(0)}_{lm}(r),w_{lm}(r)$, through which the operators
${\cal M}_{lm}^{(0)}$ and ${\cal M}_{lm}$ are defined, should have
a small distance in their functional space, the latter being
defined for two arbitrary functions $f,g$ as, for example, the
euclidian distance
\begin{equation}\label{distfunc}
d^2(f,g)=\int_{r_a}^{r_b}(f(r)-g(r))^2dr.
\end{equation}

b) Solve the eigenvalue problem (\ref{calmin2}) for the $k\times
k$ truncated problem. This determines eigenvalues $\lambda_{n'lm}$
and eigenvectors ${\mathbf d}_{n',lm}$. The latter are translated
to the new eigenfunctions $u_{n'lm}(r)={\mathbf
d}_{n'lm}^T\cdot{\mathbf u^{(0)}_{lm}(r)}$, where ${\mathbf
u^{(0)}_{lm}(r)}$ is the column matrix with the functions
$u^{(0)}_{nlm}(r), n=0,1,2,\ldots k-1$ as entries. Notice that the
matrix ${\cal M}_{lm}$ is real and symmetric, thus its
diagonalization is a numerically fast and accurate procedure.

We have implemented the above procedure in order to produce a
numerical basis set that satisfies the following two properties:

i) It is `nearby' to the HO basis set in the sense of small
functional distance of the kernel functions given by Eq.(\ref{distfunc}).

ii) It can reproduce density profiles which are shallower in the
centre than $\rho(r)\propto r^{-1}$.

The new basis set was defined as follows: Selecting the initial
kernel functions $\Phi^{(0)}_{lm}(r),\rho^{(0)}_{lm}(r)$ and basis
functions $u^{(0)}_{nlm}$ to be the HO set, we define the new
kernel functions
\begin{subequations}
\begin{align}\label{newkr} \Phi_{lm}(r)&=\Phi^{(0)}_{lm}(r)\\
\rho_{lm}(r)&=\sqrt{4\pi}\frac{1}{2\pi}\frac{(2l+1)(l+1)}{(r^2+\varepsilon^2)^{1/2}}
\frac{a r^l}{(a+r)^{2l+3}}.
\end{align}
\end{subequations}
Thus, the only change with respect to the HO set is the
introduction of a softening parameter $\varepsilon$ in the
singular factor $1/r$ of the HO density kernel function. It
follows that the operators ${\cal L}^{(0)}_{lm},{\cal L}_{lm}$ are
identical ${\cal L}^{(0)}_{lm}={\cal L}_{lm}$, but the operators
${\cal M}^{(0)}_{lm},{\cal M}_{lm}$ are different, ${\cal
M}^{(0)}_{lm}\neq {\cal M}_{lm}$, because $w^{(0)}_{lm}(r)\neq
w_{lm}(r)$. Furthermore,
$$
\lim_{\varepsilon\rightarrow 0}d(w^{(0)}_{lm},w_{lm})=0
$$
thus, for $\varepsilon$ sufficiently small, the operators ${\cal
M}^{(0)}_{lm}, {\cal M}_{lm}$ are `nearby' according to the
previously given definition.

\begin{figure}
\centerline{\includegraphics[width=8.0cm]{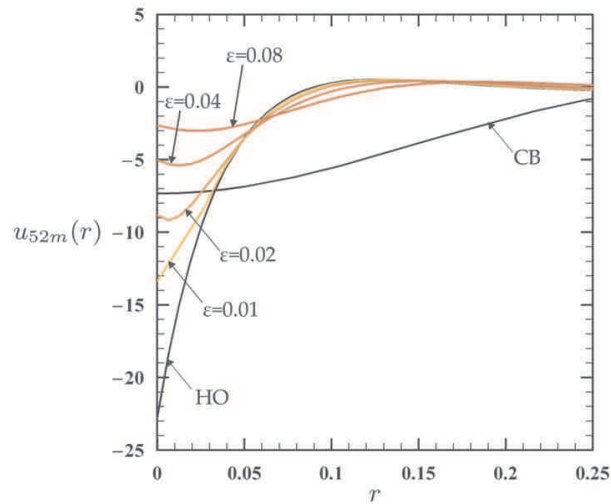}} \caption{The
forms of the radial basis functions $u_{52m}(r)$ ($n=5$, $l=2$,
independent of $m$) in the inner region ($r<0.25$) in the HO and
CB cases and in the case of the `quantum-mechanical' calculation
for various values of the parameter $\varepsilon$.}
\end{figure}

Fig.~3 shows the convergence of the numerical solution of the
eigenvalue problem (\ref{calmin2}) for the above kernel functions,
as a function of the truncation order $k$. The abscissa in
Fig.~3a,b is the dimension $k$ of the $k\times k$ matrix ${\cal
H}_{lm}^{\mbox{k-th truncation}}$ . The ordinate shows the
numerical value $\lambda_{900}$ (in the case $\varepsilon=0.02$)
of the eigenvalue of the tenth eigenvector of ${\cal
H}_{lm}^{\mbox{k-th truncation}}$ which corresponds to the quantum
numbers $n=9$, and $l=m=0$. Clearly, when $k$ is as small as
$k=11$, the numerical value of $\lambda_{900}$ calculated from the
matrix ${\cal H}_{00}^{\mbox{k-th truncation}}$ is already very
close to the value $\lambda_{900}\simeq 62.29235$ corresponding to
the limit $k\rightarrow\infty$. A zoom of Fig.~3a is shown in
Fig.~3b, in which we see that the limiting value of
$\lambda_{900}$ is practically reached after $k\gtrsim 20$. In
fact, we always find that the limiting values of the eigenvalues
from $\lambda_{0lm}$ up to $\lambda_{nlm}$ are specified
essentially up to the computer's double precision limit when one
uses matrices ${\cal H}_{lm}$ truncated at an order $k=3n$.

Fig.~4 shows an example of numerical basis function ($u_{52m}(r)$
for $n=5$, $l=2$, independent of $m$) calculated for different
values of the parameter $\varepsilon$, compared to the HO and CB
basis functions for the same quantum numbers. The plot focuses on
the region of inner radii ($r\leq 0.25$). Clearly, when $r>0.15$
all the numerical basis functions are much closer to the HO basis
set than to the CB basis set, while, at radii below $r=0.15$ the
effect of introducing $\varepsilon$ is manifested, namely a basis
function becomes less steep at the centre as the value of
$\varepsilon$ increases. Thus, by choosing different values of
$\varepsilon$ we can better control different behaviors of the
central potential or density profiles of a simulated system. We
also note that the CB basis function deviates considerably at the
centre from both the HO and the numerical basis functions, a fact
expected since the CB kernel function (Plummer sphere) represents
systems which are rather flat at the centre.

\section{A Numerical Test. The degree of order and chaos via the SCF method}

In the present section our task is to test the accuracy of
reproduction of the orbital content of a model elliptical galaxy
by various SCF codes differing in the choice of radial basis set.
The model is Denhen's (1993) $\gamma-$model with an ellipsoidal
radius $m$ (Merritt 1999, section 1). The density reads:
\begin{equation}\label{denrho}
\rho_\gamma(m) = \frac{(3-\gamma)r_a M }{4\pi
abc}m^{-\gamma}(r_a+m)^{-(4-\gamma)}
\end{equation}
with $0\leq\gamma<3$, where
\begin{equation}\label{muel}
m^2 = {x^2\over a^2}+{y^2\over b^2}+{z^2\over c^2}
\end{equation}
is the ellipsoidal radius corresponding to a triaxial system with
axial ratios $a:b:c$, $a\geq b\geq c>0$, and $M$ equal to the
total mass of the galaxy. The parameter $\gamma$ determines the
exponent of the power-law profile of the density at the centre,
i.e., $\rho(m)\sim m^{-\gamma}$ at the centre which essentially
yields also a radial profile $\rho(r)\sim r^{-\gamma}$. We are
interested in the case of `core' galaxies in which $\gamma$ is in
the range $0\leq\gamma\leq 1$. The potential corresponding to the
density (\ref{denrho}) reads (Merritt \& Fridman 1996):
\begin{equation}\label{denphi}
\Phi(x,y,z)=-\frac{M}{2(2-\gamma)r_a}
\times\int_{0}^{\infty}\tfrac{1-(3-\gamma)
\left(\tfrac{m'}{r_a+m'}\right)^{2-\gamma}+(2-\gamma)
\left(\tfrac{m'}{r_a+m'}\right)^{3-\gamma}}
{\sqrt{(\tau+a^2)(\tau+b^2)(\tau+c^2)}}d\tau
\end{equation}
with
$$
m'^2={x^2\over a^2+\tau}+{y^2\over b^2+\tau}+{z^2\over c^2+\tau}.
$$

\begin{figure}
\centerline{\includegraphics[width=15cm]{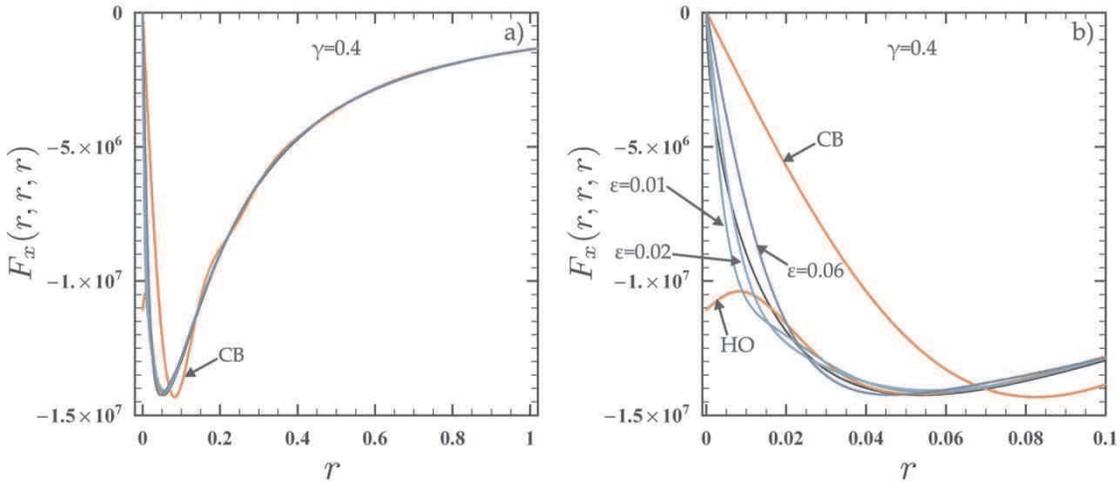}}
\caption{\textbf{(a)} The $x$ axis component $F_x$ of the force
along the direction $x=y=z=r$ for the Dehnen model $\gamma=0.4$
(black solid line) and the approximations of the same force
component obtained by implementing the SCF method to the $N$-body
realization of the Dehnen model using the HO, CB and the
$\varepsilon$-modified basis sets (the values of $\varepsilon$ are
indicated in the figure. The number of particles in the $N$-body
realization are $1.25\times 10^7$. \textbf{(b)} a focus of (a)
near the centre. The force with HO has a finite value at $r=0$,
while the force with CB is zero at the centre but its overall
central profile is much smoother than in reality. On the other
hand, all the $\varepsilon$-modified basis sets behave better than
the HO or CB sets at the centre. The optimal value of
$\varepsilon$ by visual inspection is close to
$\varepsilon=0.02$.}
\end{figure}

In order to produce an $N$-body realization of the previous
system, we work as in the numerical example of subsection (2.2)
using $N=1.25\times 10^7$ particles arranged in a $(50\times
100\times 200)$, spherical polar grid. We found that such a number
of particles was necessary because the results regarding all the
tests below were becoming robust against the number of particles
for $N$ of the order of $N=10^7$ or higher. This fact is related
to various effects caused by the Poisson noise (subsection 4.2
below). The unit of length is determined so that the half-mass
radius is at $r_{1/2}=1$, while the $N$-body system is truncated
at a radius $r=15$, containing 95\% of the total mass of the model
system (in which the distribution of the mass extends
theoretically to the limit $r\rightarrow\infty$).

In all the models we set $a=1$, $b=0.7905$, $c=0.5$ corresponding
to a maximally triaxial model. Six different Dehnen models of
progressively higher power-law exponents were examined, namely
$\gamma=0$ (perfectly harmonic core), $\gamma=0.2$, $0.4$, $0.6$,
$0.8$, and $1$ (weak cusp). After the $N$-body realization for
each of these models was produced, the potential and density were
fitted by the SCF method using eight different radial basis sets.
These are the HO and the CB sets, as well as the basis sets
derived by the `quantum-mechanical' method of section 3, for the
values of the parameter $\varepsilon$ equal to
$\varepsilon=0.005$, $0.01$, $0.02$, $0.04$, $0.06$ and $0.08$.
The latter are called $\varepsilon-$modified basis sets (modified
with respect to the HO basis set).

There are two different numerical tests performed in these
systems: a) we compare the accuracy of reproduction of the
behavior of the forces of the Dehnen model at the centre, for each
choice of SCF basis set, and b) we compute a library of 1200
orbits and compare the number of orbits that are found to be
regular or chaotic, as well as the distribution of the Lyapunov
characteristic numbers produced by the numerical integration of
the variational equations of motion in each potential
representation. In the sequel we separately analyze these two
categories of numerical tests and compare their outcome as regards
the `optimal' basis set to be used, i.e., the value of
$\varepsilon$ for which the Dehnen model and SCF agreement are
better. This information is given as a function of the value of
$\gamma$.

\begin{figure}
\centerline{\includegraphics[width=13cm]{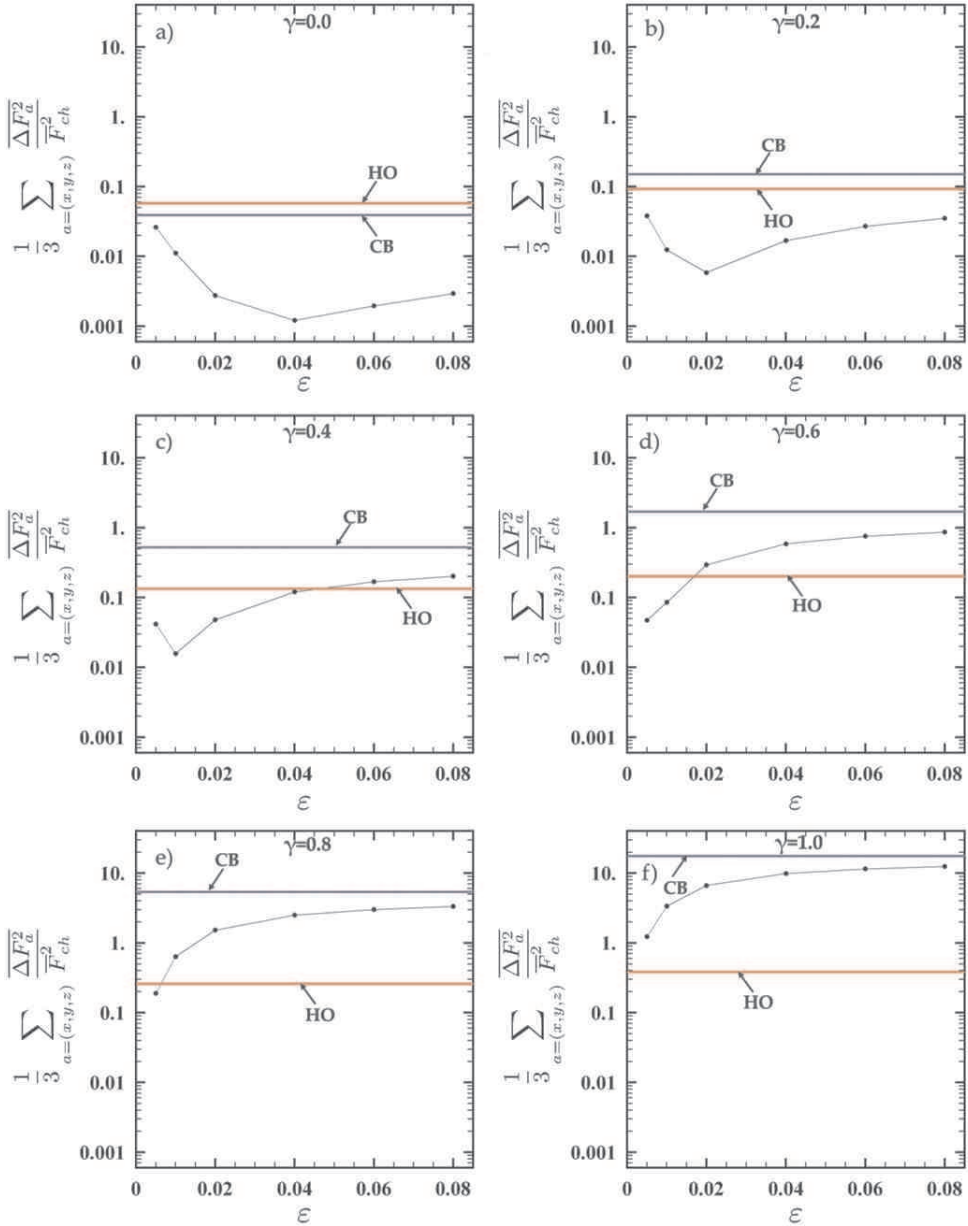}} \caption{The
quantity $\frac{1}{3}\sum\limits_{a=(x,y,z)}
\frac{\overline{\Delta F_a^2}}{F_{ch}^2}$ versus $\varepsilon$ for
each $\gamma-$model examined. The quantity $\overline{\Delta
F_a^2}$ is the mean square difference, inside a central sphere of
radius $r=0.01$, between the force component $a$ (where $a$ can be
$x, y, z$) of the specific $\gamma-$model and the same force
component as derived from different SCF basis sets implemented to
the $N$-body realization of the Dehnen model. The quantity
$\overline{F}_{ch}$ is a characteristic value of the force for all
the models set equal to $\overline{F}_{ch}=10^7$. The red and the
blue horizontal lines correspond to the HO and CB basis sets,
respectively. \textbf{(a)} $\gamma=0$, \textbf{(b)} $\gamma=0.2$,
\textbf{(c)} $\gamma=0.4$, \textbf{(d)} $\gamma=0.6$, \textbf{(e)}
$\gamma=0.8$, \textbf{(f)} $\gamma=1$. For $\gamma=0$ the minimum
error is for $\varepsilon=0.04$. As $\gamma$ increases the optimal
value of $\varepsilon$ decreases. After $\gamma=0.8$, the HO basis
set gives smaller error than the $\varepsilon-$modified basis
sets, which, however, remains in relatively higher levels than the
minimum error found for smaller values of $\gamma$. The CB basis
set gives the worst results (as regards the force error) for all
the values of $\gamma$.}
\end{figure}

\subsection{Forces}
Fig.~5 shows the reproduction of the forces by the eight different
SCF basis sets in the $N$-body realization of the $\gamma=0.4$
model, compared to the analytically derived forces through the
differentiation of (\ref{denphi}) with respect to the coordinate
variables. In all calculations below the ten first radial basis
functions $n=0,1,\ldots 9$ are used for each angular function
$Y_l^m$, with $l=0,\ldots 4$, $m=-l,\ldots,l$. We found that
despite the large increase in the number of basis functions there
was practically no significant difference observed when $n$ was
raised up to $n_{max}=13$ and $l$ up to $l_{max}=6$. The odd
angular functions $l=1,3$ or $m=1,3$ are kept in the SCF
calculation despite the fact that theoretically the coefficients
of the corresponding terms in the potential or density expansion
are zero (because the Dehnen model is fully symmetric with respect
to the three principal planes $(x-y)$, $(x-z)$, or $(y-z)$).
Numerically, these coefficients turn to have small values that
yield an estimate of the effect of the Poisson noise of the
$N$-body realization on the numerical values of the coefficients.
In Fig.~5, The $F_x$ projection of the force is plotted as a
function of $r$ when the force is calculated in the direction
$x=y=z=r$. Fig.~5a refers to a radial extent up to $r=1$, i.e., up
to the half-mass radius of the system. We see that all the SCF
models provide in general a satisfactory reproduction of the true
force in this range, except for the CB model for which the
agreement is good only in the outer radii ($r>0.2$). On the other
hand, the various models are diversified in their behavior close
to the centre, which is seen by zooming to $0\leq r\leq 0.1$ in
Fig.~5b. We notice immediately the failure of both the HO and CB
sets to provide a reasonable representation of the forces in this
scale. The force, as derived by the HO set, turns to have a
significantly non-zero value at the centre. This is precisely the
problem of the non-perfect balancing of the coefficients analyzed
in subsection (2.2). On the other hand, the CB set yields a zero
force at the centre, but near the centre the force is seriously
softened with respect to the true force. One obtains a better
agreement if one uses the $\varepsilon-$modified basis sets, as
exemplified by the plots of Fig.~5b for $\varepsilon=0.01$,
$0.02$, or $0.06$.

In order to quantify the quality of the fit of the forces by the
different SCF models, we cannot use as a relevant quantity the
fractional errors $\Delta F(r)/F(r)$, as a function of $r$,
because we have $F(r)\rightarrow 0$ as $r\rightarrow 0$, yielding
fractional errors $\Delta F(r)/F(r)\rightarrow\infty$ in this
limit. We thus use as a relevant index the quantity
\begin{equation}\label{forcerr}
<{\Delta F^2\over F^2}>={1\over 3}\sum_{\alpha=x,y,z}^3
{\overline{\Delta F_\alpha^2}\over \overline{F}_{ch}^2}
\end{equation}
where $\overline{\Delta F_\alpha^2}$ is the average value of the
squared differences of the forces in the axis $\alpha=x,y,z$ in a
spherical volume $0\leq r\leq 0.01$ and $\overline{F}_{ch}$ is a
characteristic value of the force set equal to
$\overline{F}_{ch}=10^7$, used to normalize the errors of all the
models. The quantity $<{\Delta F^2 \over F^2}>$ for the eight SCF
models, and for all the different $\gamma-$models is shown in
Fig.~6 (the case $\gamma=0.4$ of Fig.~5 corresponds to the panel
6c). In the ideal `harmonic core' case (Fig.~6a, $\gamma=0$), a
significantly non-zero value of $\varepsilon$
($\varepsilon_{optimal}\simeq 0.04$) is required in order to
minimize the error in the central forces to a level $<{\Delta
F^2\over F^2}>\sim 10^{-3}$. On the other hand, as $\gamma$
increases, the optimal value $\varepsilon_{optimal}$, at which the
force error is minimized, is shifted towards smaller values of
$\varepsilon$ ($\varepsilon_{optimal}\simeq 0.02$ when
$\gamma=0.2$ (Fig.~6b), $\varepsilon_{optimal}\simeq 0.01$ when
$\gamma=0.4$ (Fig.~6c), $\varepsilon_{optimal}\simeq 0.005$ when
$\gamma=0.6$ (Fig.~6d)). When $\gamma=1$, the
$\varepsilon-$modified basis set yields worse results than the HO
basis set, while it is still better than the CB set (Fig.~6e,f). A
typical estimate of force errors with the HO basis set is
$<{\Delta F^2\over F^2}>\sim 10^{-1}$. We emphasize that such
errors in the force determination appear inside a very small
region of the galaxy (up to $r=0.01$ when the half-mass radius is
$r=1$). Thus, one may claim that the errors are not important in
the overall $N$-body simulation. Nevertheless, these errors are
important when one calculates the regular or chaotic character of
the orbits, as demonstrated in the next subsection.

\subsection{Lyapunov exponents and the regular or chaotic character of the
orbits}

In order to check numerically the regular or chaotic character of
the orbits, we create, for each $\gamma-$model, a library of 1200
orbits calculated by a set of initial conditions uniformly
distributed on four different equipotential surfaces of the
$\gamma-$model potential $\Phi_\gamma(x,y,z)$ corresponding to the
values of the energy $E_1=0.95\Phi_\gamma(0,0,0)$,
$E_2=0.8\Phi_\gamma(0,0,0)$, $E_3=0.6\Phi_\gamma(0,0,0)$ and
$E_4=0.4\Phi_\gamma(0,0,0)$. In the energies $E_1$ and $E_2$,
which are close to the central value of the potential, we find
many `box' orbits which are regular. The existence of box orbits
is actually guaranteed by the fact that the force at the centre is
zero for $\gamma<1$. On the other hand, as the value of the energy
increases the phase space is dominated by different families of
tube orbits or chaotic orbits. The tube orbits follow essentially
the classification of Statler (1987) for the perfect ellipsoid. In
general, chaos increases as the value of $\gamma$ increases, a
fact mainly associated with the destruction of the regular
character of the box orbits.

\begin{figure}
\centerline{\includegraphics[width=8.0cm]{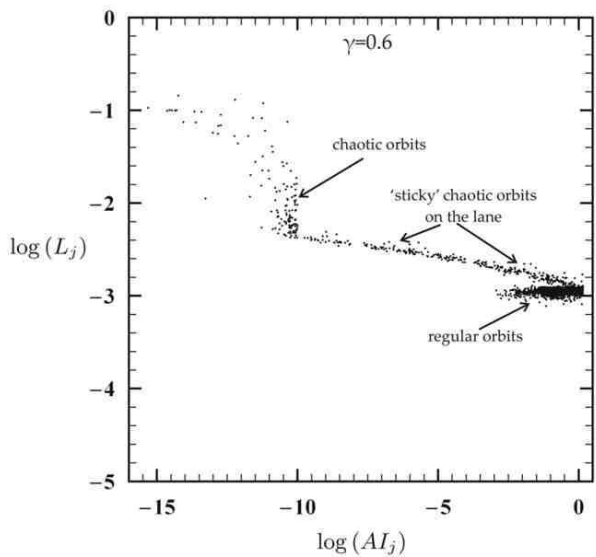}} \caption{The
method used to distinguish between regular and chaotic orbits. The
Alignment Index $AI_j$, $j=1,...,1200$ of each orbit is plotted
against the finite time Lyapunov number $L_{j}$ in $\log-\log$
scale. A detailed description of this method is given in Voglis et
al. (2002). The specific plot refers to the case of Dehnen model
with $\gamma=0.6$. Each point in this plot corresponds to an orbit
that was integrated for $1200$ radial periods. Regular orbits have
low values of $L_{j}$ (decreasing with the integration time as
$t^{-1}$) and high values of $AI_{j}$, thus the group of regular
lies in a region near $\log(L_j)\approx -3$ and $\log(AI_j)\gtrsim
-3.5$. Chaotic orbits have large values of $L_{j}$ and very small
values of $AI_j$ ($AI_{j}<10^{-10}$). The orbits on a lane joining
the above two groups are weakly chaotic orbits (i.e. sticky
orbits) which have just started exhibiting their asymptotic
chaotic behavior.}
\end{figure}

\begin{figure}
\centerline{\includegraphics[width=12cm]{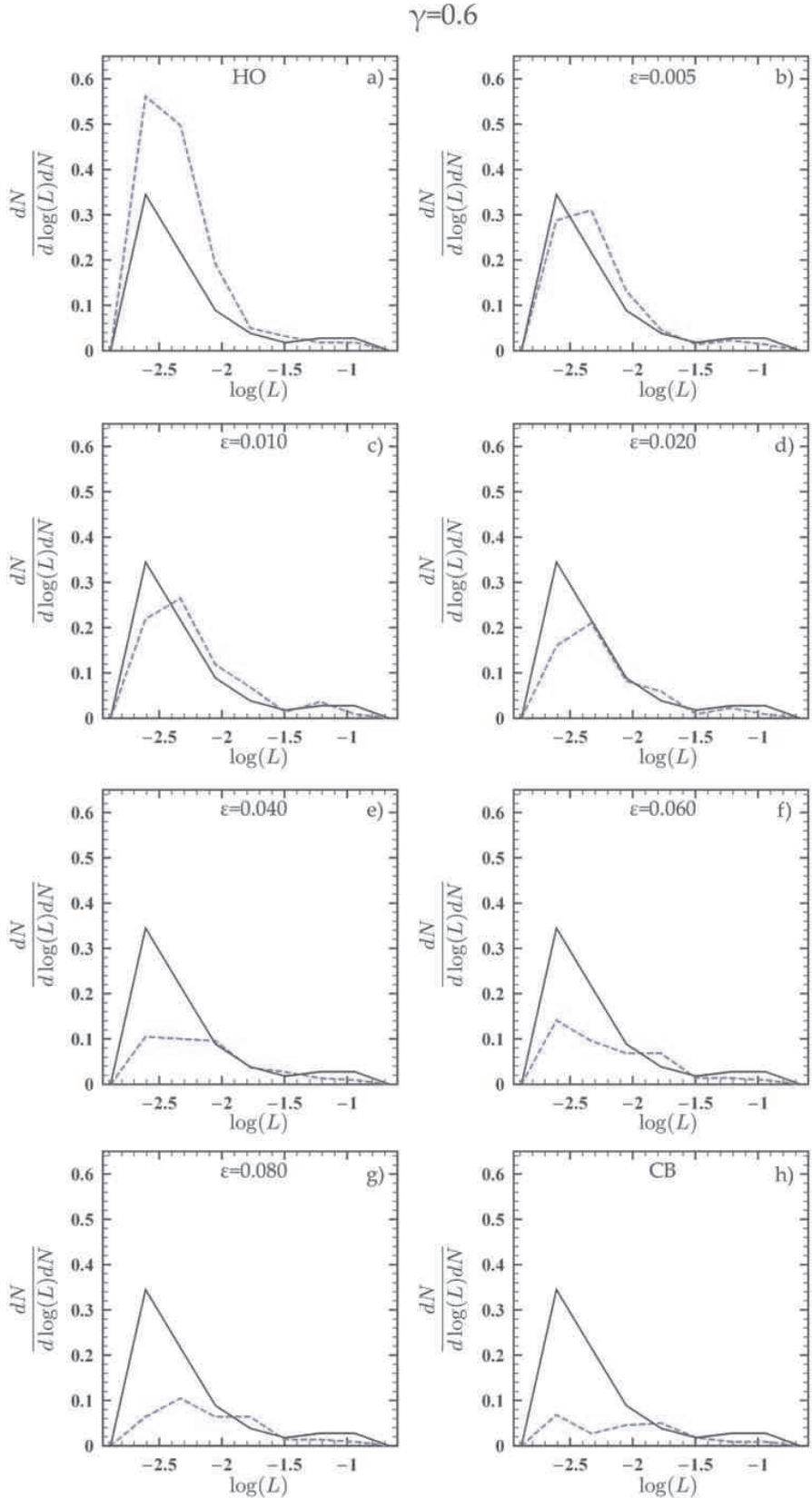}} \caption{The
distribution of the values of $\log(L)$ of the chaotic orbits for
the Dehnen model $\gamma=0.6$ (same black solid line in each
panel) compared to those found by the HO, the
$\varepsilon-$modified (for the indicated values of $\varepsilon$)
and the CB basis sets (blue dashed lines) in the $N$-body
realization of the same system. From these figures it is clear
that HO overestimates chaos, CB underestimates chaos and the best
estimation is obtained by an $\varepsilon-$modified basis set for
$\varepsilon$ between $\varepsilon=0.005$ and $\varepsilon=0.01$.
Odd angular terms $l,m=1,3$ are excluded from the computation
because they are affected by Poisson-noise asymmetries in the
$N$-body realization (see text for details).}
\end{figure}

\begin{figure}
\centerline{\includegraphics[width=16cm]{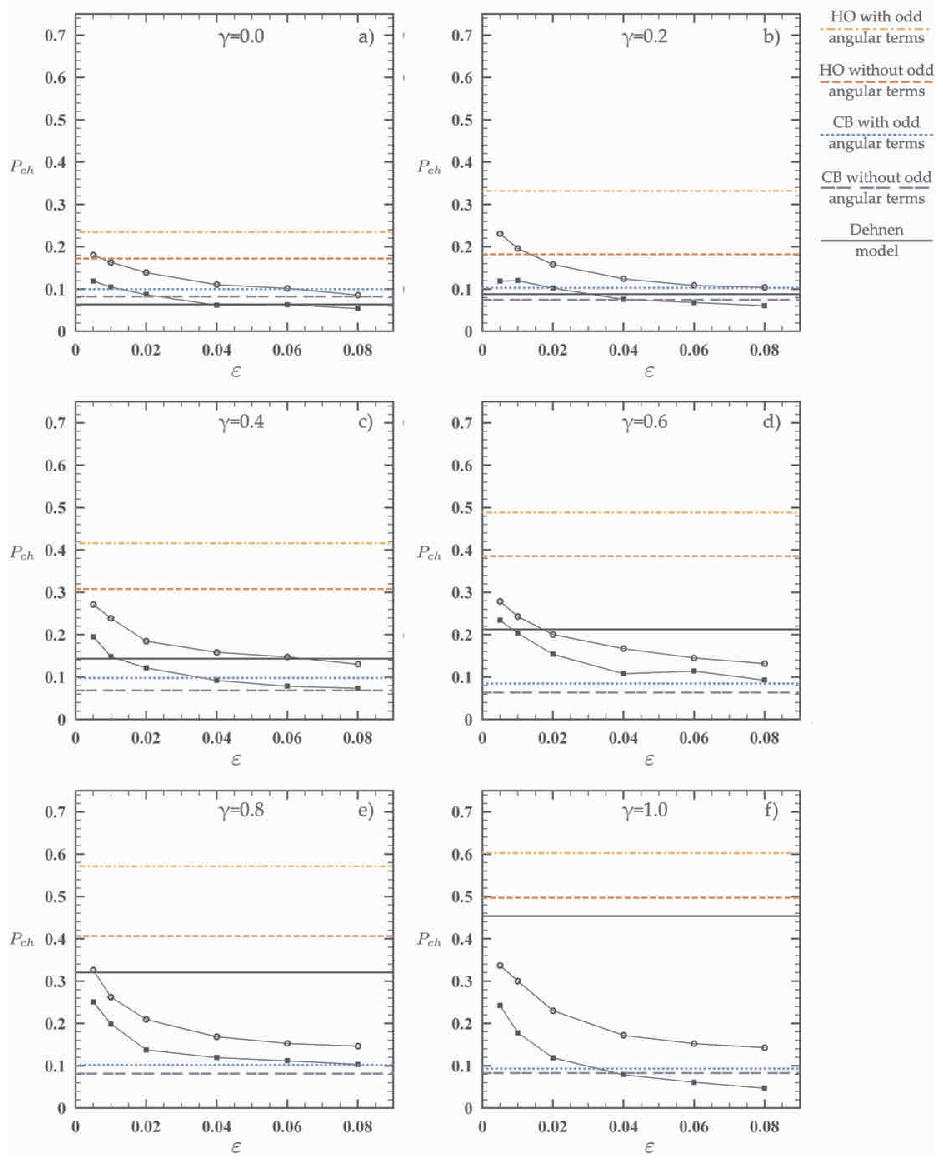}} \caption{The
percentages of chaotic orbits in the original Dehnen models and
with the various SCF basis sets applied to the corresponding
$N$-body realizations. Each panel indicates the corresponding
value of $\gamma$. Black solid lines correspond to the percentages
of chaotic orbits in the original Dehnen models, pale blue
(dotted) and blue (dashed) lines to the calculation with CB, with
and without odd angular terms, orange (dotted-dashed) and red
(dashed) lines to the calculation with HO, with and without the
odd angular terms. Black solid curves correspond to the
calculation with the $\varepsilon-$modified basis sets for the
various values of $\varepsilon$ (curves with open circles =
calculation with the full angular expansion, curves with solid
rectangles = calculation without odd angular terms). The optimal
value of $\varepsilon$ with respect to the criterion of agreement
in the percentage of chaos is determined by the point where the
curve of the $\varepsilon-$modified calculations (curve with solid
rectangles) intersects the horizontal black line (original Dehnen
model)}.
\end{figure}

In order to characterize the orbits as regular or chaotic we use
the same numerical criterion as in Voglis et al. (2002). This is a
combination of two numerical indices, called the
`finite-time-Lyapunov number' $L_j$ and the `Alignment index'
$AI_j$ respectively (see Voglis et al. 2002 for details). The
numerical values of these indices are calculated for each orbit,
labelled by the index $j$, over an integration time equal to
$T=1200$ radial periods. In a two-dimensional plot of $\log(AI_j)$
versus $\log(L_j)$ (Fig.~7, for $\gamma=0.6$), the regular orbits
accumulate in the right and middle part of the diagram, in an
almost straight horizontal segment around the value
$\log(L_j)\simeq \log(1/T)\simeq-3$. On the contrary, the chaotic
orbits occupy the up and left part of the diagram and have a
considerable scatter in the values of $L_j$, which are a measure
of the maximal Lyapunov characteristic exponent of the orbits. In
between these two groups of points there is a lane of points
connecting the two groups. This refers to weakly chaotic, or
`sticky' orbits, and as $T$ increases, there is a continuous,
albeit decreasing, flow of points through the lane towards the
group of chaotic orbits. However, this flow is irrelevant in the
characterization of the orbits for times greater than $T$, because
it should imply that some orbits characterized as `chaotic' have,
in fact, Lyapunov times much longer than the age of the galaxy.

Fig.~8 shows, now, the main result as regards the choice of an
appropriate set of basis functions that better reproduces the
regular or chaotic character of the orbits. All the panels show
the distributions of the logarithm of the finite-time Lyapunov
numbers $L_j$ in the case $\gamma=0.6$ for the orbits which are
within or above the transport lane of Fig.~7, i.e., the orbits
characterized as chaotic. The distributions are {\it not}
normalized, i.e., the area below each distribution is equal to the
percentage of orbits that were characterized as chaotic. Thus,
differences in the total area covered by two distributions mean a
difference in the total percentage of the orbits characterized as
chaotic, while differences in the shape reflect a different
normalized distribution of Lyapunov exponents. The distribution
shown with solid line refers to the precise calculation in the
analytic Dehnen force field, while dashed plots refer to
computations with different SCF basis sets. Furthermore, in this
figure all the SCF distributions refer to a calculation in which
we switched off the odd terms of the angular expansion of the
potential as derived by the SCF code for reasons explained
immediately below.

\begin{figure}
\centerline{\includegraphics[width=7cm]{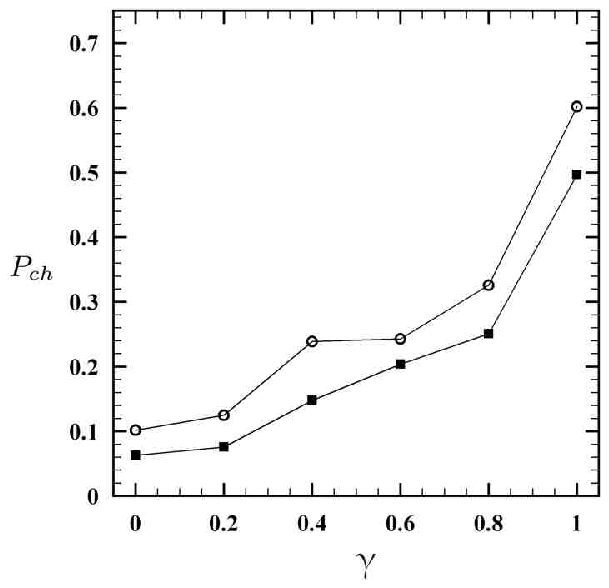}} \caption{The
percentages of chaotic orbits found by the use of the optimal
basis set (see Fig.~9), for each value of $\gamma$, when the
non-symmetric (odd) angular terms in the potential expansion are
turned on (lower curve) or off (upper curve).}
\end{figure}

\begin{figure}
\centerline{\includegraphics[width=12cm]{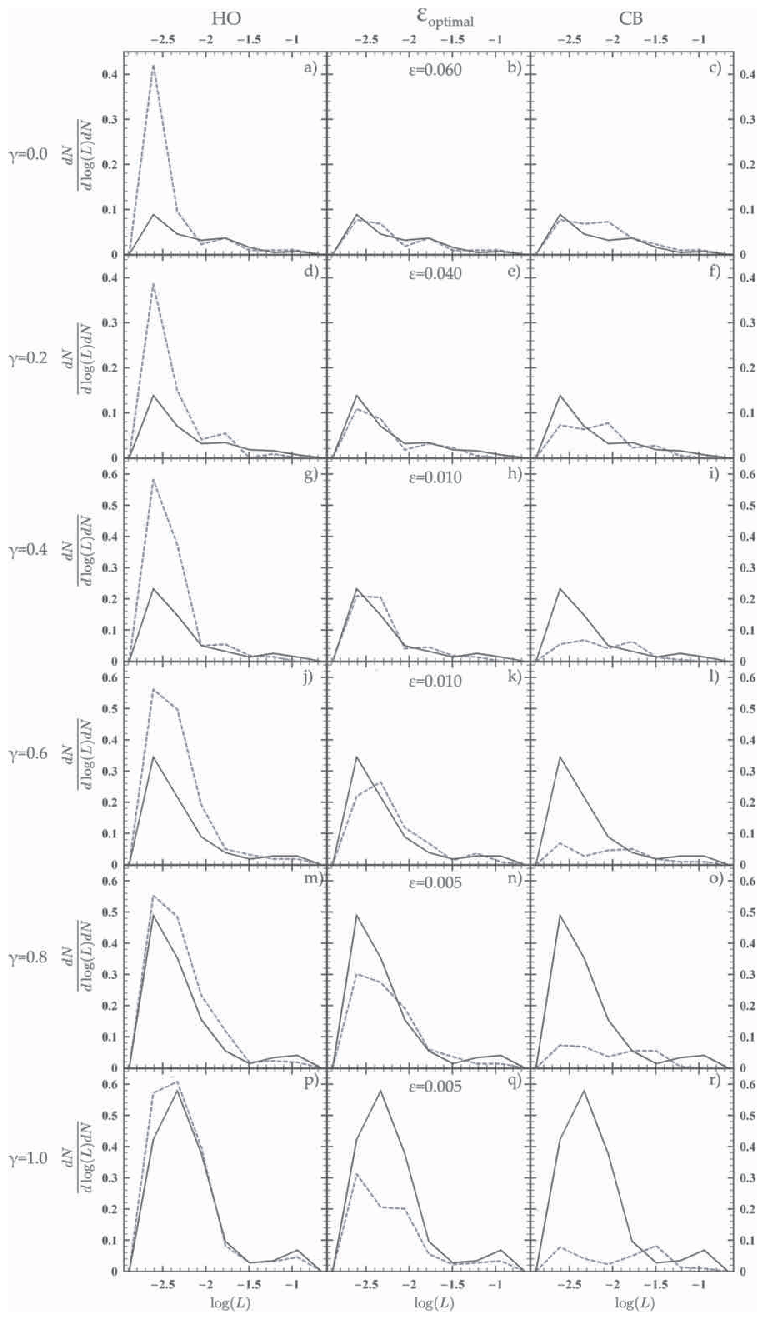}} \caption{The
$\log(L)$ distributions of the chaotic orbits as derived by the
HO, optimal $\varepsilon-$modified, and CB approximations compared
to the true distribution of the corresponding Dehnen model. Each
row refers to one indicated value of $\gamma$. Black solid curves
correspond to the real distribution while blue dashed curves are
the distributions with HO (left column), optimal
$\varepsilon-$modified (middle column), or CB (right column) we
have for the specific value of $\gamma$. For $\gamma\leq 0.6$ the
agreement is better between the real and optimal
$\varepsilon-$modified calculations. For $\gamma=0.8$ the results
with $\varepsilon-$modified ($\varepsilon=0.005$) and HO
($\varepsilon=0$) are comparable, while for $\gamma=1.0$ the best
result is with HO.}
\end{figure}

The following are some basic remarks regarding these diagrams:

a) The computation with the HO basis set overestimates by a factor
of about 1.7 the percentage of chaotic orbits in this model. Thus,
as shown below, while the true percentage for $\gamma=0.6$ is
$\simeq 22\%$, the HO fit yields $\simeq 38\%$. In fact, the
difference in the two percentages is even higher if one takes into
account the odd terms of the angular expansion of the potential
(in this case we find a percentage $48\%$ with the HO basis set).
However, we know that the true value of these terms in the Dehnen
model is equal to zero, so that non-zero values can only be
associated with asymmetries in the number of particles on the two
sides of any of the principal planes of the galaxy induced by the
Poisson noise of the $N$-body realization. The simplest way to
measure the latter is by measuring the size of the coefficients of
the odd angular functions $l=1,3$ or $m=1,3$ in the potential
expansion. We thus use the following measure of the Poisson noise:
\begin{equation}\label{oddcoef}
PN = {||\mbox{coefficients of odd angular terms}||\over||\mbox{all
coefficients}||}
\end{equation}
where the norm $||\cdot||$ means sum of the absolute values of the
coefficients. In the case of Fig.~8a, we find $PN\simeq 2.3\times
10^{-3}$ which is consistent with an estimate $PN\sim
O(1/\sqrt{N})$ with $N=1.25\times 10^7$. This looks like a
relatively small number, which however produced a 10\% difference
in the percentage of chaos with the HO simulation.

b) The computation with the CB set (Fig.~8h) underestimates the
percentage of chaotic orbits by a factor $\simeq 2.75$ (22\% true
percentage against 8\% with CB). The underestimate is even more
serious (6\%) if one switches off the odd angular terms. The forms
of the two distributions are also quite different, a fact meaning
that the CB fit is rather unsuitable to represent a galaxy with
central power-law exponent $\gamma=0.6$ (or beyond).

c) The best results are found when we use an
$\varepsilon-$modified basis set with $\varepsilon_{optimal}$
between the values $\varepsilon_{optimal}=0.005$ and
$\varepsilon_{optimal}=0.01$ (Fig.~8b,c). For this value of
$\varepsilon$ both the percentage of chaotic orbits and the
distribution of the Lyapunov exponents derived by the SCF method
are in good agreement with the true percentage and distribution.

Fig.~9 shows a comparison of the true percentages of chaotic
orbits with those found by the SCF fit with different basis sets,
for all the examined values of $\gamma$. An optimal value of
$\varepsilon$, based on the `percentage of chaos' criterion, is
determined by the point where the curves with open dots intersect
the horizontal solid line, marking the true percentage. This
value, however, is also contaminated by chaos due to the Poisson
noise on the odd angular terms, and a better determination is made
when these terms are switched off (solid curves with black
rectangles). In any case, Fig.~9 renders immediately clear that
the HO fit produces overestimates of the percentage of chaotic
orbits in the whole range of values $0\leq\gamma\leq 1$, while the
CB fit underestimates the percentage of chaos for the values
$0.4\leq\gamma$. Up to $\gamma = 0.8$ an optimal
$\varepsilon-$modified model can be found yielding the best
agreement with the true percentage. Near this value of $\gamma$,
however, the situation is reversed, and in the limit
$\gamma\rightarrow 1$ it is the $\varepsilon-$modified models
yielding underestimates and the HO model yielding the best
results. This is, precisely what is expected by noticing the
$r^{-1}$ power-law profile of the HO monopole terms at the centre.

Fig.~10 shows the comparison of the percentages of chaotic orbits
found by the use of the optimal basis set, for each value of
$\gamma$, when the non-symmetric (odd) angular terms in the
potential expansion are turned on or off. In all cases, the
numerical noise due to the non-symmetric (odd) terms increases the
percentage of chaotic orbits. This fact justifies the use of
symmetrized N-Body realizations of a system, as e.g. by
Holley-Bockelmann et al. (2001), when an orbital analysis is
requested.

Fig.~11 shows a comparison of the distributions of the finite-time
Lyapunov exponents of the chaotic orbits for all the $\gamma-$models
considered, with the distributions derived via the
HO model, optimal $\varepsilon-$modified, and CB model. This
figure checks essentially whether, at the value of $\varepsilon$
at which the agreement of the percentages of chaotic orbits is
good, the agreement of the distributions of the finite time
Lyapunov numbers is also good. This check is necessary, because it
is possible that two very different distributions yet cover the
same total area. The plots in the middle column of Fig.~11 clearly
show that the distributions with the $\varepsilon-$modified basis
set, at the optimal value $\varepsilon=\varepsilon_{optimal}$, are
indeed in good agreement with the true distributions, except in
the case $\gamma=1$ in which the best agreement is obtained by the
HO basis set.

\begin{figure}
\centerline{\includegraphics[width=12.5cm]{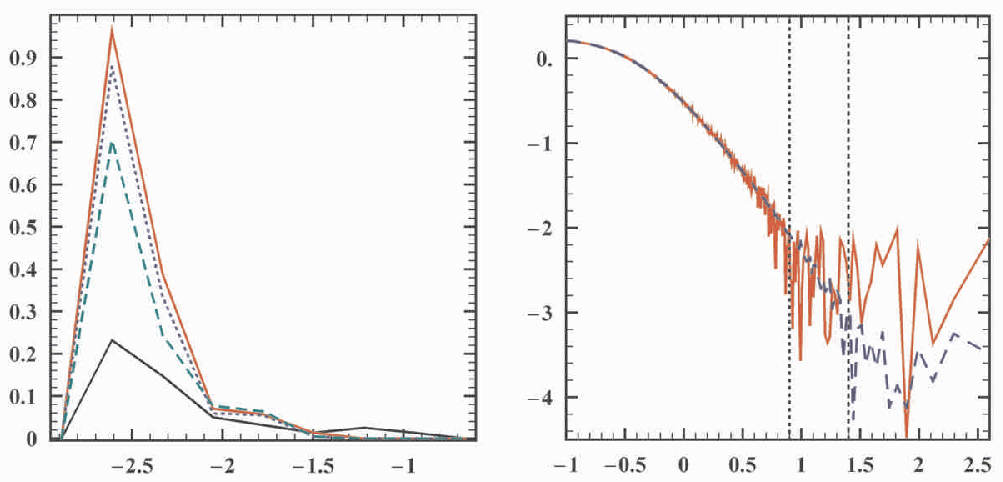}}
\caption{\textbf{(a)} The $\log(L)$ distributions of the chaotic
orbits, as derived by the HO basis set (with all terms), in a
rejection-acceptance Monte Carlo realization of the $\gamma=0.4$
Dehnen model, when the truncation radius is $r_t=15$ (red solid
line), $r_t=30$ (blue dotted line), $r_t=100$ (green dashed line).
The real distribution by the original $\gamma$-model is plotted by
black solid line. \textbf{(b)} The absolute error
$|I(r_t)-I(\infty)|$ versus $r_t$ in $\log-\log$ scale. The red
solid line corresponds to evaluations of $I(r_t)$ using Monte
Carlo realizations with $N=10^5$ particles while the blue dashed
line to Monte Carlo realizations with $N=10^7$ particles. We see
that beyond a value of $r_t$ the error due to Poisson noise
dominates the error due to finite radius $r_t$.}
\end{figure}

\begin{figure}
\centerline{\includegraphics[width=12.5cm]{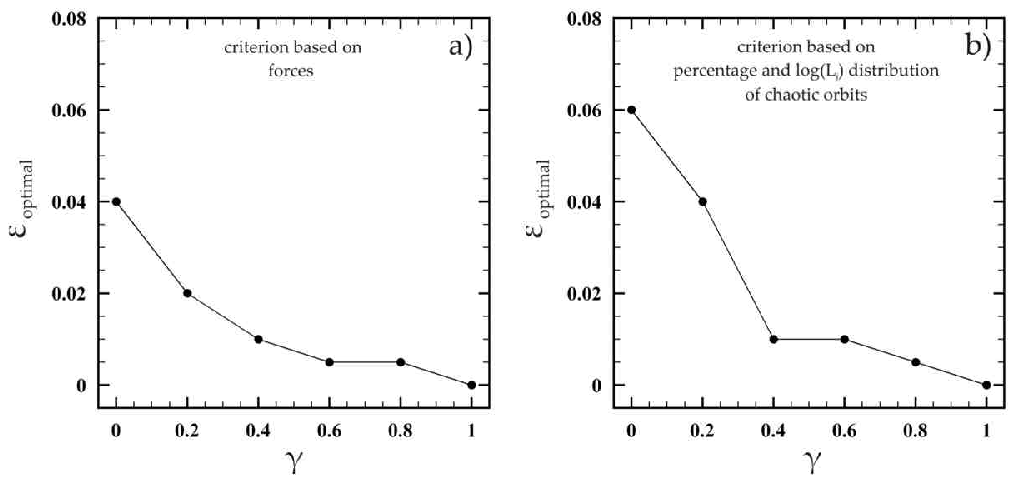}}
\caption{\textbf{(a)} The optimal value of $\varepsilon$ versus
$\gamma$ based on the `forces' criterion (see Fig.~6).
\textbf{(b)} the optimal value of $\varepsilon$ versus $\gamma$
based on the criterion of the percentage and $\log(L)$
distribution of the chaotic orbits (see Figs.~9 and 10).}
\end{figure}

In order to check the robustness of the above results on the
particular way chosen to create the N-Body realization of a Dehnen
$\gamma-$model, we created a different realization, in which there
is no use of polar grid, but an acceptance-rejection Monte Carlo
algorithm was utilized to produce an N-Body system for the
$\gamma=0.4$ model. The mass contained in a small volume element
corresponding to values of the spatial coordinates in the
intervals $(m,m+dm)$, $(\theta, \theta+d\theta)$, and
$(\phi,\phi+d\phi)$ is given by $dm =
m^2abc\rho(m)\sin(\theta)dm~d\theta~d\phi$, implying that the mass
distribution is separable and uniform in the coordinates
$X_1=m^3/3$, $X_2=-\cos(\theta)$ and $X_3=\phi$. We thus readily
obtain acceptance-rejection Monte Carlo realizations of the
models. In practice we transform $m$ to $\xi=(m-1)/(m+1)$ so that
when $\xi$ is given values in $-1\leq\xi< 1$, $m$ can take
arbitrarily large values. In reality, given the finite number of
particles in the simulation, the distribution of the matter is
undersampled at large radii. We thus checked the robustness of our
results against three different truncation radii, namely a)
$r_t=15$ (which is the standard choice in the previous
simulations), b) $r_t=30$, and c) $r_t=100$. The results are
summarized in Fig.~12a, for the distributions of the finite-time
Lyapunov numbers calculated with the HO basis set. When the
truncation radius is at $r_t=30$, there is a marginal improvement
of the results (the total percentage of chaotic orbits found is
37\%, against 41\% when $r_t=15$, and 14\% real percentage). But
even with a much higher truncation radius ($r_t=100$) the
improvement is still small (32\% against 14\% real percentage of
chaotic orbits). We thus conclude that the distributions shown in
Fig.~11 do not change appreciably by changing the sampling
technique or the truncation radius.

A simple argument can show why the scaling of the numerical error
with the truncation radius $r_t$ saturates at some value of $r_t$.
This is based on equating the error in the evaluation of
generalized integrals, due to a truncation of the limits of
integration, with the error due to the Poisson noise at large
radii by a Monte Carlo evaluation of the integral. Namely, since
the Monte Carlo sampling uses a finite number of particles, the
drop of the density at large radii implies that the mass
distribution is seriously undersampled at these radii. An example
is given in Fig.~12b, which refers to a calculation  of the
integral
\begin{equation}\label{imc}
I(r_t)=\int_0^{r_t}4\pi r^2\rho(r)\Phi_{00}(r)dr,
\end{equation}
corresponding essentially to a truncation at finite radius of the
generalized integral (\ref{bnml1}) for a spherical Dehnen-model
$\rho(r) =r^{-\gamma}(1+r)^{-(4-\gamma)}$ when the oscillatory
behavior of the functions $u_{n00}(r)$ is neglected. The integral
(\ref{imc}) was evaluated by a Monte Carlo sum using $N=10^5$ or
$N=10^7$ particles. Fig.~12b shows the absolute error
$|I(r_t)-I(\infty)|$ when a different Monte Carlo realization of
the system is produced for each value of $r_t$. For small $r_t$,
the error is large and it is due to the truncation of the
generalized integral at insufficiently small radius. However,
beyond a value of $r_t$, the Poisson noise due to the
undersampling of the density at the outer parts clearly dominates,
causing variations of the error which are of the same order as the
error due to the finite truncation.  When $N=10^5$, the critical
value of $r_t$ is estimated as $r_t=8$ (in units of the half-mass
radius), and the error beyond this radius fluctuates between
$10^{-2}$ and $10^{-3}$. On the other hand, for $N=10^7$, the
truncation radius beyond which the Poisson noise dominates raises
to $r_t=25$, and the error beyond this radius fluctuates between
$10^{-4}$ and $10^{-3}$. These values are compatible with the
estimates given independently in section 2.

Finally, Fig.~13 shows a comparison of the optimal value of
$\varepsilon$ determined in the previous analysis by (a) the
central force criterion (subsection 3.1), and (b) the chaotic
percentage criterion (present subsection). The latter criterion
suggests the use of somewhat larger values of $\varepsilon$ than
by the former criterion, when $\gamma< 0.4$, while the two
criteria yield nearly equal values for $\gamma\geq 0.4$. At any
rate, the main conclusion drawn from the above numerical examples
is that the characterization of the regular or chaotic character
of the orbits of an $N$-body system depends crucially on the
accurate numerical representation of the `smooth' potential that
presumably underlies the mass distribution of the $N$ particles.
In the framework of the SCF method, this conclusion is translated
into the need for very careful choice of basis set. In the above
examples we used a `quantum-mechanical' method in order to
calculate numerical basis sets which are close to the HO basis
set, but they can better fit the behavior of all smooth quantities
(potential, forces, density) at the centre when the power-law
central density profile is shallower than $r^{-1}$. Such profiles
naturally arise in $N$-body simulations of the remnants of galaxy
mergers (e.g. Jesseit et al. 2005 and references there in).
Nevertheless, the method is applicable, with different starting
basis sets, to a much wider class of stellar dynamical systems and
the cautions raised in the present section are, very probably,
relevant to such systems as well.

\section{Conclusions}

In the present paper we address the question of the choice of an
optimal basis set of functions for orbital studies of galaxies
simulated via the self-consistent field (SCF) method. Our approach
is to create $N$-body realizations of a Dehnen analytical
$\gamma-$model of a triaxial galaxy and then reproduce the
gravitational potential by the SCF method, using different basis
sets which are either analytical. i.e., the Hernquist-Ostriker and
Clutton-Brock sets, or numerical, depending on a parameter
$\varepsilon$ that modifies the behavior of the potential at the
centre with respect to the HO basis set. We then compare a number
of quantities characterizing the orbits (central force profiles,
percentage of chaotic versus regular orbits and distributions of
the Lyapunov characteristic numbers) in the original model and in
the SCF reproduction of the potential. Our main conclusions are
the following:

1) If the kernel functions (section 2) used in the construction of
a radial basis set have a singular behavior at the centre, this
basis set is unsuitable for simulating galaxies which do not
exhibit the same singular behavior. In particular, even if the
singularity can be theoretically removed by a balancing of the
coefficients of some terms in the SCF series, the numerical noise
induced by the Monte-Carlo evaluation of the values of the
coefficients destroys the balance and results in large errors
appearing in the evaluation of the forces or derivatives of the
forces, especially in the central parts of the galaxy.

2) When following the methodology suggested by Weinberg (1999) for the
determination of numerical radial basis sets, shooting methods yielding
tabulated values of the basis functions on a grid should be avoided,
because they, too, introduce large errors in the evaluation of the
derivatives of the forces, yielding large inaccuracies in the integration
of the variational equations of motion.

3) We propose a `quantum-mechanical' method of determination of numerical
radial basis sets that overcomes the above problems. This method is based
on the diagonalization of a truncated matrix which results from a spectral
formulation of the Sturm-Liouville problem, equivalent to the procedure
referred to in quantum mechanics as the diagonalization of the Hamiltonian
matrix. The stability and accuracy of this scheme is demonstrated by
specific numerical examples.

4) We study the orbits in a family of triaxial Dehnen
$\gamma-$models for various values of $\gamma$ in the range $0\leq
\gamma\leq 1$, corresponding to the case of elliptical galaxies
with `shallow' central cusps. For each $\gamma-$model we use a
family of different radial basis functions which are modifications
of the HO basis set, derived by implementing the
`quantum-mechanical' algorithm of (3). The basis functions depend
on one small parameter $\varepsilon$ defined so as to yield the HO
basis set in the limit $\varepsilon\rightarrow 0$. The numerical
tests concern comparisons of a) the behavior of forces in the
central parts of the galaxy, b) the percentage of regular and
chaotic orbits, and c) the distribution of the Lyapunov exponents
of the chaotic orbits, between the exact $\gamma-$model and the
various $N$-body - SCF realizations using the HO, CB and
$\varepsilon-$modified models for various values of $\varepsilon$.
When the standard basis sets (HO or CB) are used, we find large
deviations in all three criteria between the exact and SCF
results, except for the HO basis set for $\gamma=1$. On the other
hand, the best agreement is found by using $\varepsilon-$modified
basis sets for values of $\varepsilon$ in the range $0.005\leq
\varepsilon \leq 0.06$ (in units of the half mass radius). The
optimal value of $\varepsilon$ is given as a function of the value
of $\gamma$. The latter can in principle be determined in advance
(before or during the $N$-body simulation) by measuring the
power-law exponent of the central radial profile of the system.

All the numerical codes used in the present paper are available by
the authors upon request.

\section*{Acknowledgments}

Stimulating discussions with Professor G. Contopoulos are
gratefully acknowledged. This research was supported in part by
the Research Committee of the Academy of Athens. C. Kalapotharakos
was supported in part by the Empirikion Foundation, and C.
Efthymiopoulos by an Archimidis - EPEAEK grant of the Ministry of
National Education.

\label{lastpage}

\end{document}